\documentclass[aps,prb,reprint,floatfix,showpacs,superscriptaddress]{revtex4-1}
\usepackage{txfonts}
\usepackage{graphicx}
\usepackage[colorlinks,citecolor=magenta,linkcolor=blue]{hyperref}

\begin{document}


\title{A ``hidden'' transition in heavy fermion compound CeB$_{6}$ at $T\sim$ 20 K?}
\author{Haiyan Lu}
\email{hylu@iphy.ac.cn}
\affiliation{Science and Technology on Surface Physics and Chemistry Laboratory, P.O. Box 9-35, Jiangyou 621908, China}
\affiliation{Beijing National Laboratory for Condensed Matter Physics, and Institute of Physics, Chinese Academy of Sciences, Beijing 100190, China}
\author{Li Huang}
\email[Corresponding author: ]{lihuang.dmft@gmail.com}
\affiliation{Science and Technology on Surface Physics and Chemistry Laboratory, P.O. Box 9-35, Jiangyou 621908, China}
\date{\today}

\begin{abstract}
The temperature-dependent electronic structures of heavy fermion compound CeB$_{6}$ are investigated thoroughly by means of the combination of density functional theory and single-site dynamical mean-field theory. The band structure, density of states, and 4$f$ valence state fluctuation of CeB$_{6}$ are calculated in a broad temperature range of 10 $\sim$ 120\ K. Overall, the 4$f$ electrons remain incoherent, approximately irrespective of the temperature. However, we find that these observables exhibit some unusual features near 20\ K. In addition, the evolution of 4$f$ orbital occupancy, total angular momentum, and total energy with respect to temperature shows apparent singularity around 20\ K. We believe that these tantalizing characteristics are probably related to a ``hidden'' electronic transition.    
\end{abstract}


\maketitle


\section{introduction\label{sec:intro}}

The Ce-based heavy fermion compounds, in which the Ce-4$f$ valence electrons and the $c-f$ hybridization effect play essential roles, have attracted extensive research interests in recent years~\cite{PhysRevLett.86.684, PhysRevLett.100.176402}. There exists complex and subtle interplay between the spin, orbital, and lattice degrees of freedom, which leads to a rich physics of these systems, including heavy fermion superconductivity~\cite{RevModPhys.56.755, RevModPhys.63.239, PhysRevLett.92.027003, PhysRevLett.87.057002}, quantum criticality~\cite{PhysRevLett.117.016601, PhysRevLett.116.206401, PhysRevB.92.155147}, Kondo effect~\cite{PhysRevLett.99.136401}, and mixed-valence behavior~\cite{PhysRevLett.102.026403, PhysRevB.36.5739}, to name a few. Many exotic ordered phases emerge as a result of the interaction between the strong electronic correlations, large spin-orbit coupling, and intricate crystal-field splitting in these materials, which make them good testing beds for exploring new quantum phenomena and physical mechanisms.

Numerous theoretical and experimental investigations suggested that the electronic and magnetic structures of Ce-based heavy fermion compounds are very sensitive to the change of external environment, such as temperature, pressure, chemical doping, and electromagnetic field~\cite{yuan:2016}. Here we focus on the temperature effect only. The temperature-driven evolution of electronic and magnetic structures has already been observed in many Ce-based heavy fermion materials. For example, the magnetic susceptibility and resistivity measurements for Kondo semiconductor CeNiSn demonstrate that a pseudogap arises when $T < 12$ K~\cite{PhysRevB.41.9607}. The Ce-4$f$ spectral function of CeCoGe$_2$ develops a ``kink" owing to the Kondo resonance and the formation of hybridization gap when temperature is reduced~\cite{PhysRevB.88.125111}. Perhaps the most representative examples are the Ce-based ``115" compounds (Ce$M$In$_{5}$, $M$ = Co, Rh, and Ir). For instance, the Ce-4$f$ electrons in CeIrIn$_5$ evolve from localized to itinerant states upon cooling, accompanied by a remarkable change of Fermi surface from small to large volume~\cite{PhysRevLett.108.016402}. The topology and volume of Fermi surface of CeCoIn$_5$ are also tuned by temperature significantly~\cite{PhysRevB.90.125147}. They are still subjects of ongoing researches.

Now let us turn to cerium hexaboride, a typical Kondo dense system with Kondo temperature $T_{\text{K}}\approx 1$\ K. CeB$_6$ crystallizes in a simple cubic structure with space group $Pm\bar{3}m$. The Ce ions locate at the cubic center, while the boron octahedra situate at the cubic corners. The experimental lattice parameter $a_0$ is 4.1396 {\AA} determined by single crystal diffraction~\cite{BLOMBERG1991313}. As a paradigm of heavy fermion compound, the specific heat coefficient $\gamma$ of CeB$_{6}$ is about 250 mJ $\cdot$ mol$^{-1}$ $\cdot$ K$^{-2}$, which is much larger than the one of LaB$_{6}$~\cite{jpcm7823}. One of the most fascinating properties of CeB$_{6}$ is its low-temperature magnetic structures and the corresponding magnetic phase transitions, which have been studied for more than half a century. Under ambient pressure and when $T>$10 K, CeB$_6$ is paramagnetic. Below $T_{\text{Q}}$=3.3 K, an antiferro-quadrupolar order forms, while the conventional dipolar antiferromagnetic order develops below $T_{\text{N}}$=2.4 K~\cite{EFFANTIN1985145}. Many efforts have been devoted to figure out the magnetic phase diagram of CeB$_{6}$~\cite{KAWAKAMI1980435,JPSJ.52.728,PhysRevB.69.054415,Kobayashi281,PhysRevB.69.014423,Sluchanko2007,BRANDT1985937,EFFANTIN1985145,ERKELENS198761,Horn1981}. Though great progresses have been achieved, until now there are many puzzles that need to be solved. For example, the magnetic ordered phases revealed by neutron scattering~\cite{Horn1981,EFFANTIN1985145,ERKELENS198761,PhysRevLett.86.1578,PhysRevB.68.214401,nmat3976} are not consistent with those by nuclear magnetic resonance~\cite{JPSJ.52.728}. In addition, the appearance of a ``magnetically hidden order" phase has been confirmed by X-ray diffraction experiment~\cite{PhysRevLett.103.017203}, but the long suspected multipolar origin of this phase is still under intense debate nowadays. The correlated electronic structure of CeB$_{6}$ is yet another important, but less concerned topic. The band structure and Fermi surface topology have been studied by using the high-resolution angle-resolved photoemission spectroscopy (ARPES)~\cite{PhysRevB.92.104420}. However, the temperature dependence of electronic structures of CeB$_{6}$, especially the evolution of 4$f$ electronic states, is rarely considered in the available literatures. From the theoretical side, since the electronic correlations among Ce-4$f$ electrons are strong which are hardly handled by the classic band theory, and the cubic symmetry of CeB$_{6}$ enables an active orbital degree of freedom for Ce-4$f$ valence electrons, even a reasonable picture for its electronic structure is still lacking. Hence, to study the temperature-tuned electronic structures of CeB$_{6}$ by \emph{ab initio} calculations becomes an unprecedented task up to this time. 

In order to fill in this gap, we endeavored to uncover the temperature-dependent electronic structures of CeB$_6$ by using a first-principles many-body approach, namely the density functional theory in combination with the single-site dynamical mean-field theory (dubbed as DFT + DMFT)~\cite{RevModPhys.78.865,RevModPhys.68.13}. We successfully reproduced the bulk properties and ARPES experimental results of CeB$_{6}$ at first. And then we found that its electronic structures didn't evolve monotonically with respect to temperature as expected. They and the corresponding physical observables exhibited sudden changes at $T \sim 20$ K. Thus, we postulate that a ``hidden" electronic transition may occur around this temperature.  

The rest of this paper is organized as follows: Section~\ref{sec:method} briefly introduces the technical details for the DFT + DMFT calculations. In Sec.~\ref{sec:results}, the temperature-dependent electronic structures, including the momentum-resolved spectral function, 4$f$ valence state fluctuation, 4$f$ electronic configuration, 4$f$ orbital occupancy, total angular momentum, and total energy are presented. Several possible mechanisms for this ``hidden" transition are discussed and analyzed in Sec.~\ref{sec:disc}. Finally, a concise summary is addressed in Sec.~\ref{sec:summary}.  

\section{method\label{sec:method}}

The DFT + DMFT method, which combines the first-principles aspect of DFT with the non-perturbative many-body treatment of local interaction effects in DMFT, may be the most powerful established approach to study the strongly correlated systems~\cite{RevModPhys.78.865,RevModPhys.68.13}. It has been successfully applied to investigate the physical properties of many Ce-based heavy fermion materials in recent years~\cite{shim:1615,PhysRevB.81.195107,PhysRevLett.108.016402,PhysRevB.94.075132,PhysRevB.88.125111}. In view of the correlated feature of the Ce-4$f$ states in CeB$_{6}$, we adopted the DFT + DMFT method to perform charge fully self-consistent calculations to explore the fine electronic structures of CeB$_{6}$ as a function of temperature.

We used the \texttt{WIEN2K} code~\cite{wien2k} and the \texttt{DMFT\_W2K} package~\cite{PhysRevB.81.195107} to carry out the DFT and DMFT calculations, respectively. In the DFT part, $R_{\text{MT}}K_{\text{MAX}} = 7.0$, the $k$-points mesh was $17 \times 17 \times 17$, and the generalized gradient approximation (Perdew-Burke-Ernzerhof functional)~\cite{PhysRevLett.77.3865} was employed to express the exchange-correlation potential. In the DMFT part, the general interaction matrix was parameterized using the Coulomb interaction $\mathcal{U}$ and the Hund's exchange $\mathcal{J}$ via the Slater integrals~\cite{PhysRevB.59.9903}. They were 6.0 eV and 0.7 eV, respectively, which were exactly accordance with the values reported in the literatures~\cite{PhysRevB.92.104420}. The fully localized limit scheme was used~\cite{jpcm:1997} to evaluate the double counting term. The constructed multi-orbital Anderson impurity models were solved using the hybridization expansion continuous-time quantum Monte Carlo impurity solver (dubbed as CT-HYB)~\cite{RevModPhys.83.349,PhysRevLett.97.076405,PhysRevB.75.155113}. In order to accelerate the calculations, we adopted the the trick of lazy trace evaluation~\cite{PhysRevB.90.075149}. In addition, we not only utilized the good quantum numbers $N$ (total occupancy) and $J$ (total angular momentum) to classify the atomic eigenstates, but also made a truncation ($N \in$ [0,3]) for the local Hilbert space~\cite{PhysRevB.75.155113} to reduce the computational burden.
 
All of the calculations were carried out using the experimental crystal structure~\cite{BLOMBERG1991313}. The temperature range we considered was from 10 K to 120 K, so it was reasonable to retain only the paramagnetic solutions. The convergence criteria for charge and energy were $10^{-4}$ e and $10^{-4}$ Ry, respectively. The final output were Matsubara self-energy function $\Sigma(i\omega_n)$ and impurity Green's function $G(i\omega_n)$, which were then utilized to obtain the integral spectral function $A(\omega)$ and momentum-resolved spectral function $A(\mathbf{k},\omega)$. From the probability of atomic eigenstates, we can extract key information concerning the 4$f$ valence state fluctuation and electronic configuration.  

\section{results\label{sec:results}}

\subsection{Benchmark calculations\label{subsec:bench}} 

\begin{figure}[t]
\centering
\includegraphics[width=\columnwidth]{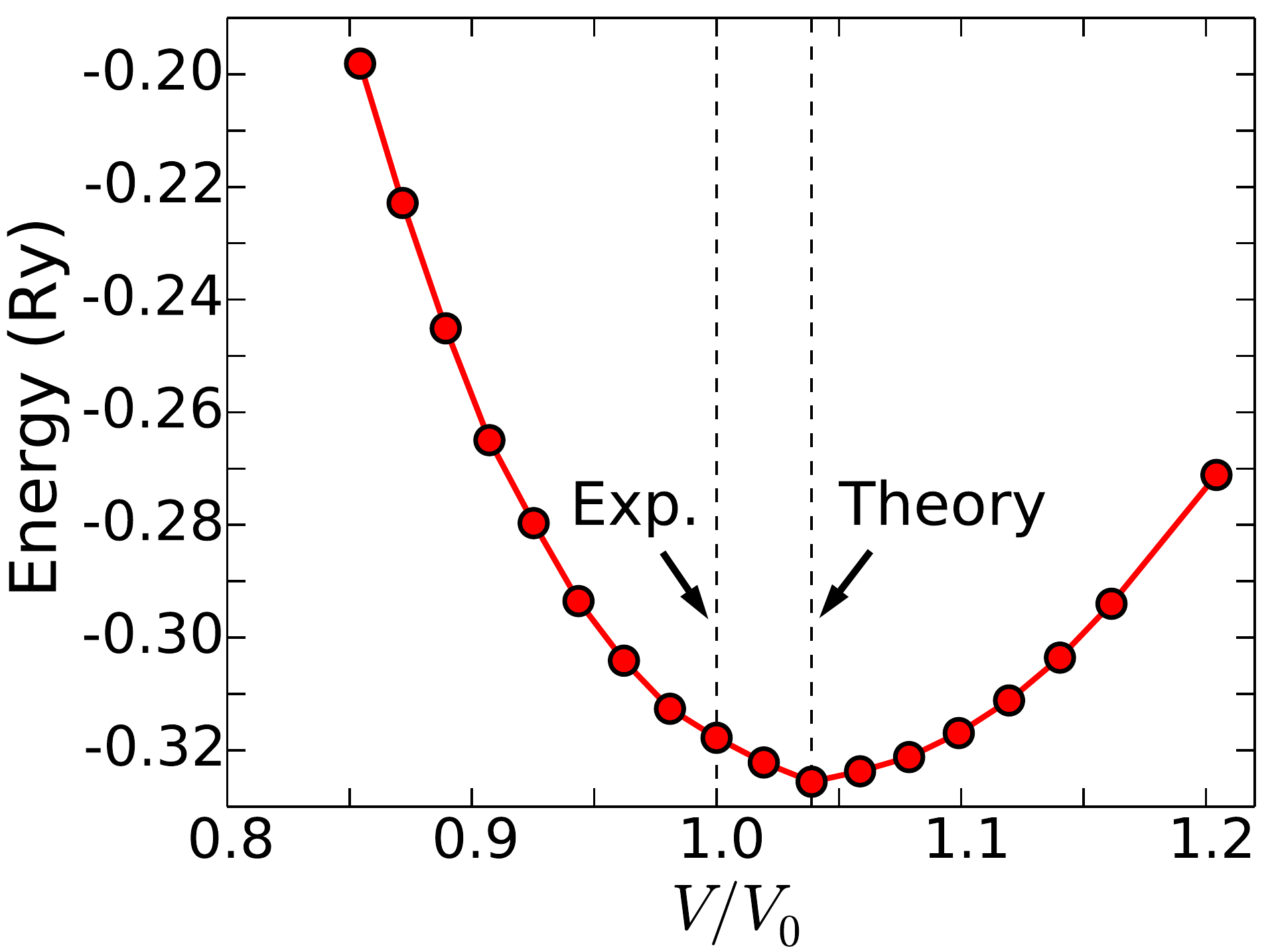}
\caption{(Color online). Calculated $E-V$ curve of CeB$_{6}$. A reference energy (-18029 Ry) is subtracted from the total energy data. Here $V_{0}$ denotes the experimental crystal volume~\cite{BLOMBERG1991313}. The two vertical dashed lines indicate the experimental and theoretical equilibrium crystal volumes ($V/V_0 = 1.00$ and $V/V_0 \sim 1.04$), respectively. \label{fig:exp1}}
\end{figure}

\begin{figure}[t]
\centering
\includegraphics[width=\columnwidth]{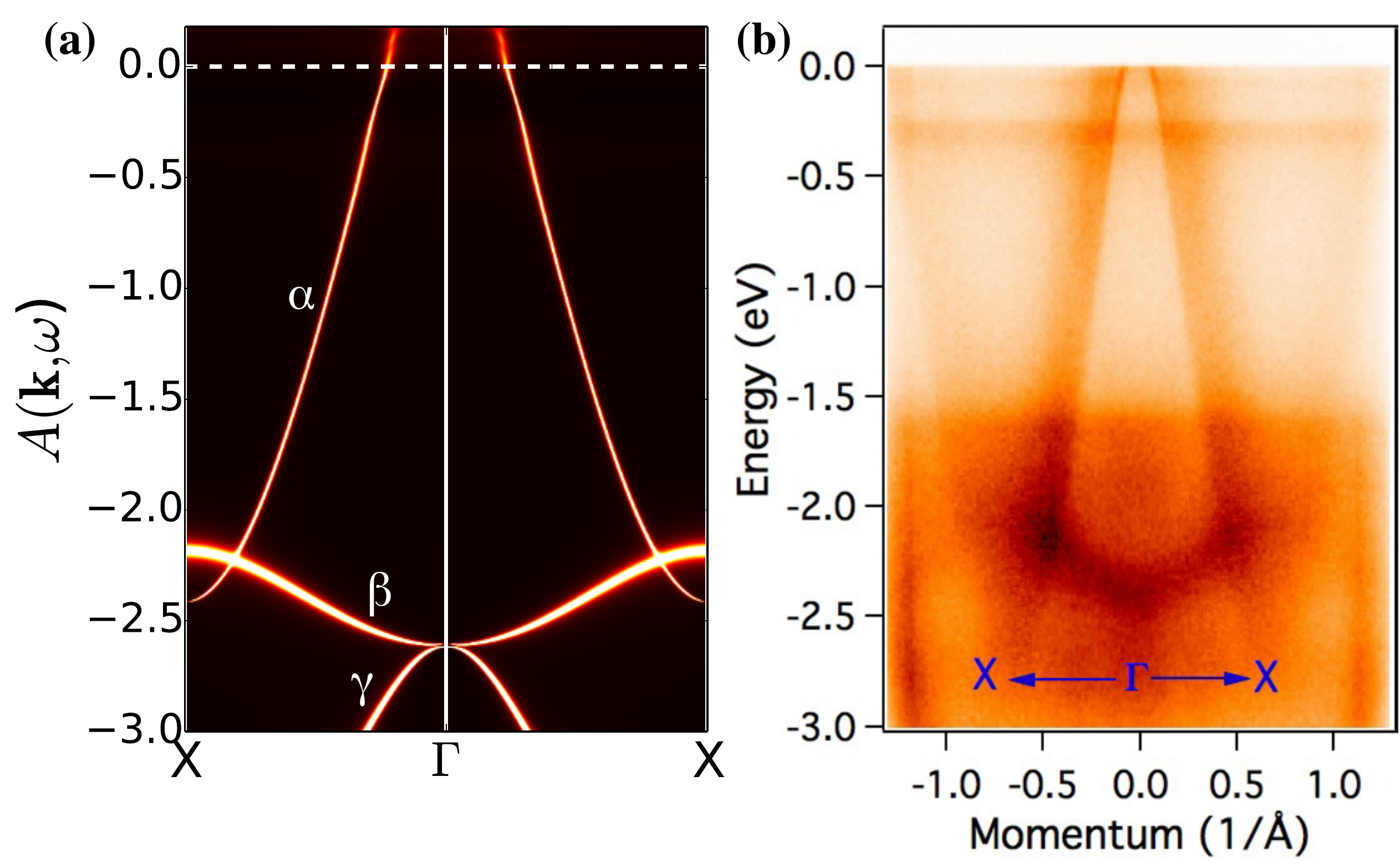}
\caption{(Color online). (a) The momentum-resolved spectral function $A(\mathbf{k},\omega)$ obtained by DFT + DMFT calculations under ambient pressure at $T = 16.6$ K. The horizontal dashed line means the Fermi level. (b) Band dispersions measured by ARPES at $T = 17$ K. Reproduced from Reference~[\onlinecite{PhysRevB.92.104420}]. \label{fig:exp2}}
\end{figure}

\begin{table}[t]
\caption{Bulk properties of CeB$_{6}$. Here $B$ and $a_0$ denote the bulk modulus (GPa) and equilibrium lattice constant (\AA), respectively. \label{tab:bulk}}
\begin{ruledtabular}
\begin{tabular}{ccc}
Methods & $B$ & $a_0$ \\
\hline
DFT + DMFT  & 169 & 4.2031\footnotemark[1] \\
LDA         & 173 & 4.1542\footnotemark[2] \\
Experiments & 191\footnotemark[3], 166\footnotemark[4], 168\footnotemark[5], 182\footnotemark[6] & 4.1396\footnotemark[7]
\end{tabular}
\end{ruledtabular}
\footnotetext[1]{The DFT + DMFT calculations were performed at $T = 11.6$ K.}
\footnotetext[2]{See Reference~[\onlinecite{PhysRevB.82.104302}].}
\footnotetext[3]{See Reference~[\onlinecite{JPSJ.63.623}].}
\footnotetext[4]{See Reference~[\onlinecite{LEGER1985995}].}
\footnotetext[5]{Measured by ultrasonic method at room temperature. See Reference~[\onlinecite{Lüthi1984}].}
\footnotetext[6]{Measured by Brillouin scattering technology at room temperature. See Reference~[\onlinecite{Lüthi1984}].}
\footnotetext[7]{Measured at $T = 298$ K. See Reference~[\onlinecite{SATO1985310}].}
\end{table}

In order to examine the correctness and reliability of the DFT + DMFT method, we applied it to study the bulk properties of CeB$_{6}$ at first. We calculated the $E-V$ curve (see Fig.~\ref{fig:exp1}), and then used the Birch-Murnaghan equation of states (EOS) to fit it. The extracted bulk modulus ($B$) and equilibrium lattice parameter ($a_0$), together with the available experimental and theoretical data, are collected and displayed in Tab.~\ref{tab:bulk}. Clearly, our calculated results are in good consistent with the experimental values. The error is less than 4.5\%

Next we tried to calculate the momentum-resolved spectral function $A(\mathbf{k},\omega)$ of CeB$_{6}$ along the high-symmetry lines in the Brillouin zone ($T \sim 16.6$ K). Then the obtained $A(\mathbf{k},\omega)$ is compared with the band dispersion measured by ARPES experiment at $T = 17$ K~\cite{PhysRevB.92.104420}. In the calculated $A(\mathbf{k},\omega)$ we easily identify a hole-type Fermi surface, a quasi-linear band ($\alpha$), and two parabolic-like bands ($\beta$ and $\gamma$) near the $\Gamma$ point, which are also observed in the experimental spectrum (see Fig.~\ref{fig:exp2}). Note that we also performed a regular DFT calculation to compute the band structure. But it fails to reproduce the ARPES results. It is apparent that the neglect of the 4$f$ electronic correlation should be responsible for this failure. 

From the results presented above (bulk properties and band structures), we may arrive at the conclusion that the DFT + DMFT method is a reliable tool to describe the electronic states of CeB$_6$, and could be used for further calculations. 

\subsection{Temperature-dependent electronic structures}

\begin{figure*}[t]
\centering
\includegraphics[width=\textwidth]{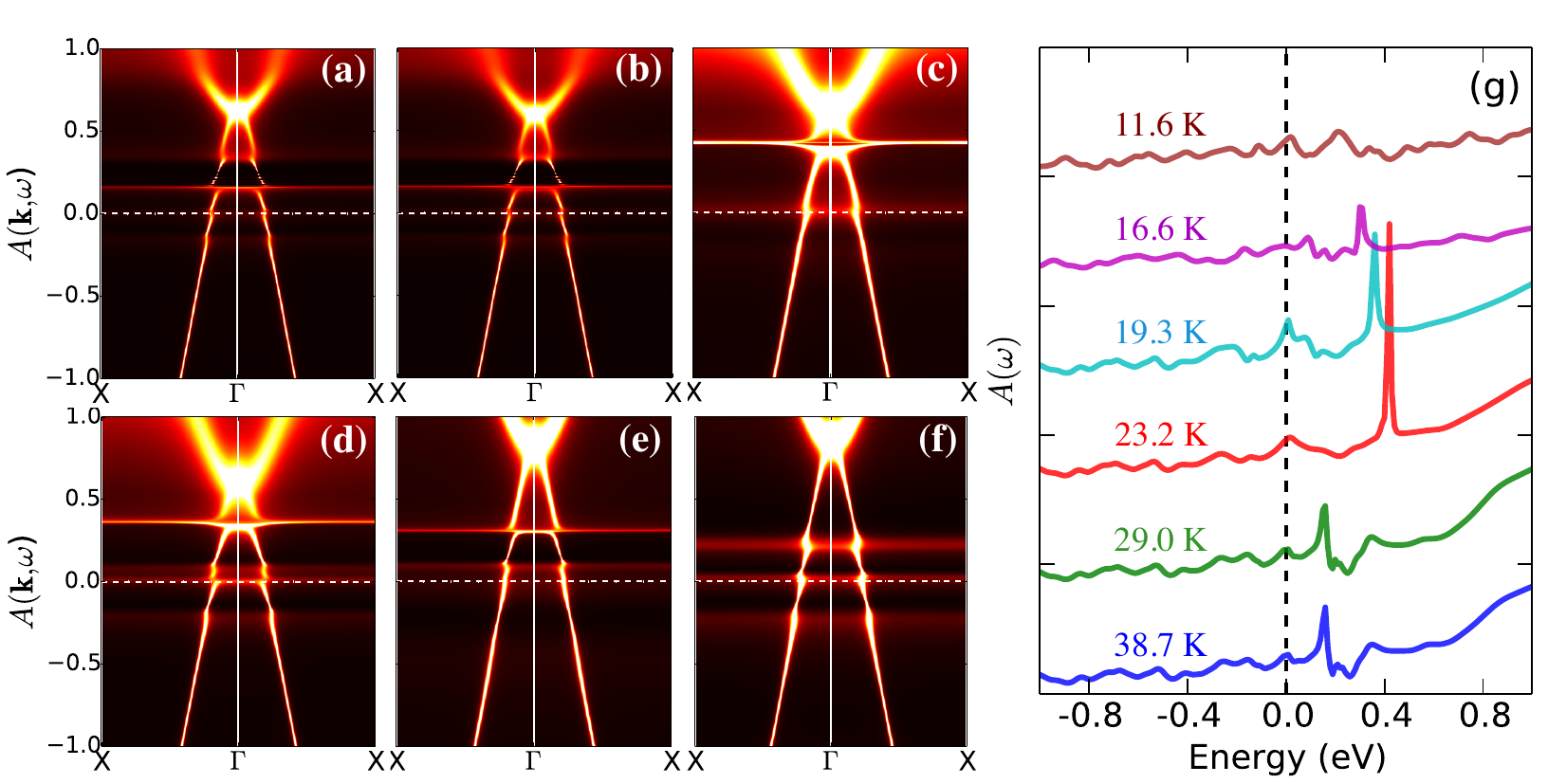}
\caption{(Color online). Evolution of momentum-resolved spectral functions $A(\mathbf{k},\omega)$ upon temperature obtained by DFT + DMFT calculations under ambient pressure. (a) 38.7 K. (b) 29.0 K. (c) 23.2 K. (d) 19.3 K. (e) 16.6 K. (f) 11.6 K. The corresponding integrated spectral functions $A(\omega)$ (density of states) are shown in panel (g). The horizontal or vertical dashed lines mean the Fermi level. \label{fig:akw}} 
\end{figure*}

\begin{figure*}[t]
\centering
\includegraphics[width=\textwidth]{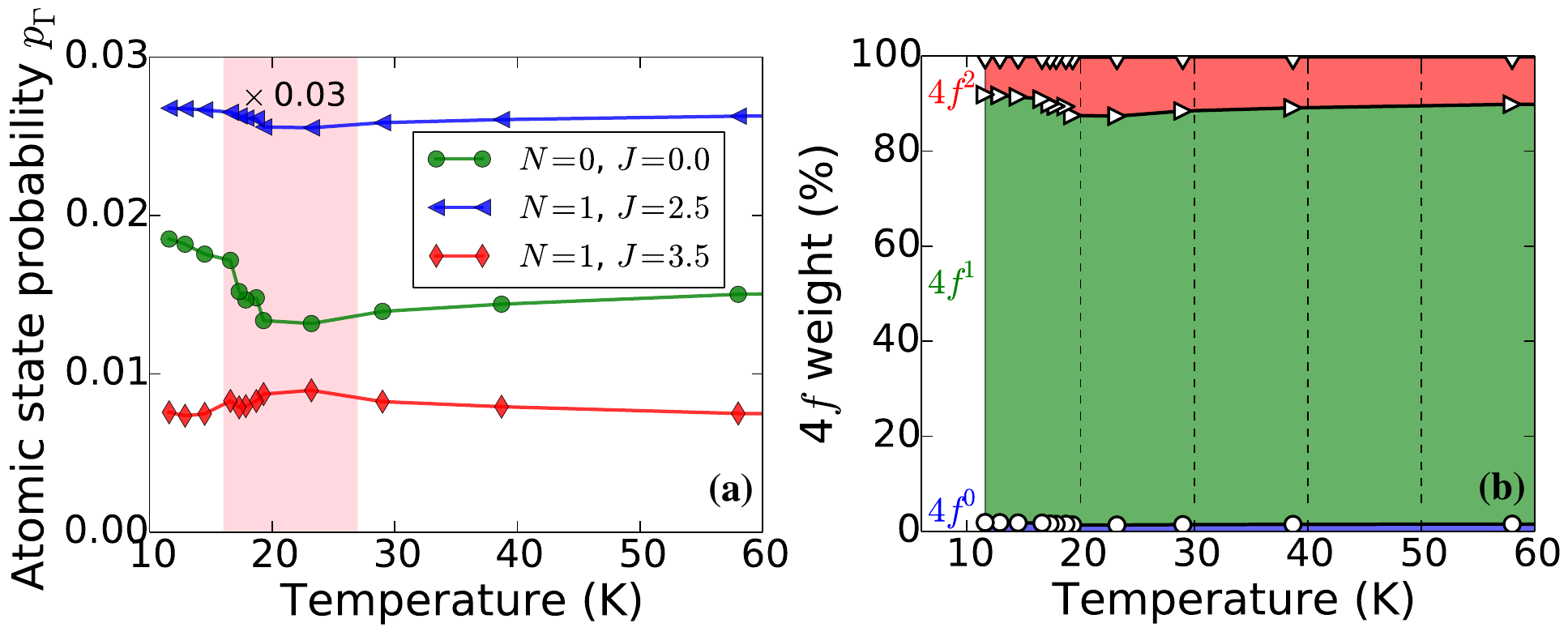}
\caption{(Color online). (a) Evolution of atomic eigenstates probabilities $p_\Gamma$ upon temperature obtained by DFT + DMFT calculations under ambient pressure. Here the data for the $|N = 1, J = 2.5\rangle$ atomic eigenstates are rescaled (multiplied by a factor of 0.03) for a better visualization. The vertical color bar denotes the possible transition zone. (b) Ce-4$f$ electronic configurations as a function of temperature. Note that the percentages for the $4f^{3}$ electronic configurations ($N = 3$) are too small ($< 1\%$) to be seen. At $T \sim 20$ K, the $4f^1$ electronic configuration ($N = 1$) has the smallest percentage. \label{fig:prob1}}
\end{figure*}

\begin{figure*}[t]
\centering
\includegraphics[width=\textwidth]{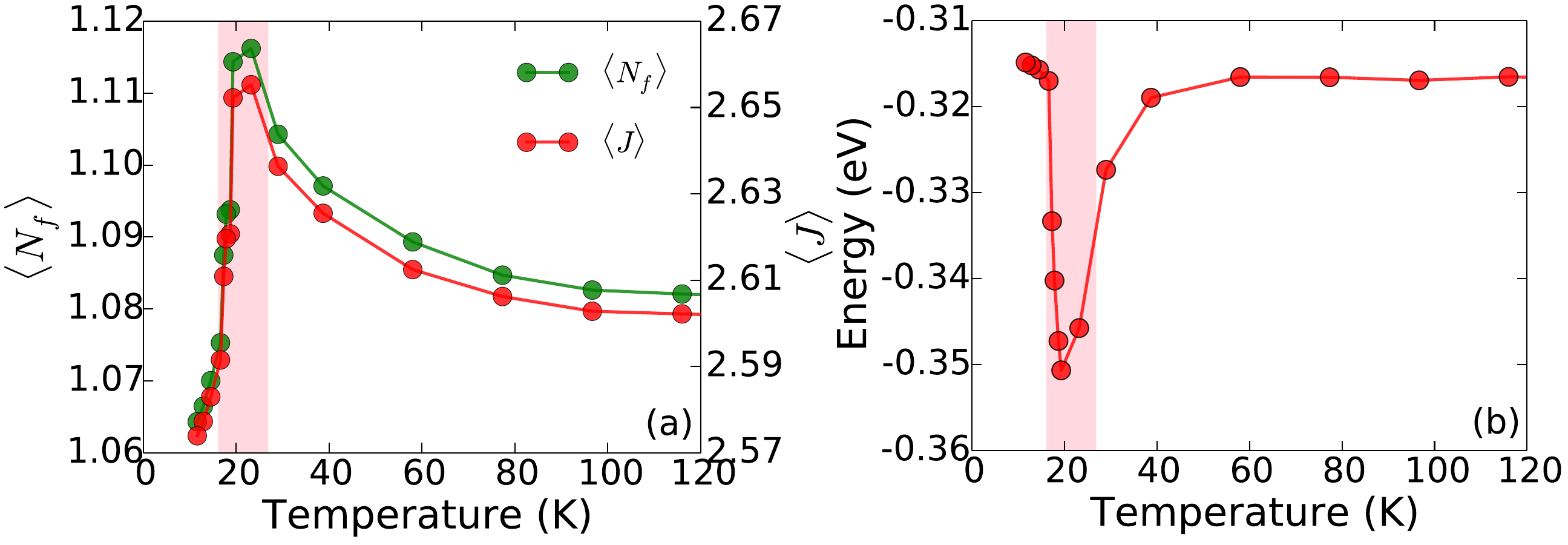}
\caption{(Color online). Evolution of some physical observables upon temperature obtained by DFT + DMFT calculations under ambient pressure. The possible transition zone is marked by vertical color bar. (a) Expected values of 4$f$ occupancy $\langle N_{f} \rangle$ (left $y$-axis) and total angular momentum $\langle J \rangle$ (right $y$-axis). (b) Total energy $E_{\text{DFT + DMFT}}$. A reference energy (-18029 Ry) was already subtracted from the original data for a better visualization. \label{fig:prob2}}      
\end{figure*}

\emph{$T$-dependent band structures.} In Fig.~\ref{fig:akw}, the momentum-resolved spectral functions $A(\mathbf{k},\omega)$ for selected temperatures at the $X - \Gamma - X$ lines in the Brillouin zone are illustrated. The most prominent feature shown in $A(\mathbf{k},\omega)$ is the flat bands which belong to the 4$f$ orbitals. Because of the presence of strong spin-orbit coupling, the 4$f$ bands are split into $j = 5/2$ and $j = 7/2$ sub-bands. The energy separation is approximately 300\ meV, which is consistent with experimental observation $\Delta_{\text{SO}}$=280 meV~\cite{PhysRevB.92.104420}. In addition, the 4$f$ orbitals are almost unoccupied. Thus, the $j = 5/2$ (low-lying) and $j = 7/2$ (high-lying) sub-bands are parallel and above the Fermi level. When they cross the conducting bands, there is strong $c-f$ hybridization. All of these features are very similar to those observed in Ce metal and the other Ce-based heavy fermion compounds, such as CeIn$_{3}$~\cite{PhysRevB.94.075132} and CeCoIn$_{5}$~\cite{PhysRevB.81.195107}. 

When the temperature is lowering, the momentum-resolved spectra show considerable changes. The overall trend is that the Ce-4$f$'s $j = 5/2$ and $j = 7/2$ sub-bands are shifted toward the Fermi level gradually, while the energy difference between the two sub-bands remain almost unchanged. However, we observe some surprising things when $T \sim 20$ K [see Fig.~\ref{fig:akw}(c) and (d) for $T = 23.2$ K and $T = 19.3$ K, respectively]. The $j = 5/2$ sub-bands are close to the Fermi level, and become dimmer. On the other hand, the $j = 7/2$ sub-bands are pushed back to higher energy, and their intensities are largely enhanced. Of course, the energy difference between them increases obviously. These changes seen in $A(\mathbf{k},\omega)$ are also simultaneously represented by the integrated spectral functions $A(\omega)$ [i.e, density of states, see Fig.~\ref{fig:akw}(g)]. The temperature-driven evolution of $A(\omega)$ can be divided into two separate stages. (i) From $T \sim 40$ K to $ T \sim 20$ K, the peak near 0.4 eV which is corresponding to the $j = 7/2$ sub-bands, shifts to the higher energy and becomes much stronger. The splitting induced by spin-orbit coupling turns larger. (ii) From $T \sim 20$ K to $T \sim 10$ K, the change is completely reversed. The spin-orbit splitting is reduced. To sum up, we find that the band structure and density of states of CeB$_{6}$ show unexpected changes when $T \sim 20$ K, which is probably related to some kind of transitions or crossovers.  

There is still one more thing we would like to emphasize. The 4$f$ spectral weights near the Fermi level are mainly from the contributions of the $j = 5/2$ components. They are a bit small and not very sensitive to the change of temperature. In other words, the 4$f$ electrons in CeB$_{6}$ are highly localized, and virtually don't get involved into the chemical bonding. No localized to itinerant transition~\cite{shim:1615} is observed in the temperature range studied in the present work. 

\emph{$T$-dependent valence state fluctuations and electronic configurations.} Here, we turn to the 4$f$ valence state fluctuations and electronic configurations of CeB$_{6}$ upon temperature. The CT-HYB quantum impurity solver is capable of computing the valence state histogram for 4$f$ electrons, which denotes the probability to find out a 4$f$ valence electron in a given atomic eigenstate $|\Gamma\rangle$ (labelled by good quantum numbers $N$ and $J$ as mentioned at Sec.~\ref{sec:method})~\cite{PhysRevB.75.155113}. In Fig.~\ref{fig:prob1}(a), the probabilities for the most important three atomic eigenstates as a function of temperature are plotted. As a first glimpse, we find that the $|N = 1, J = 2.5\rangle$ atomic eigenstate is overwhelmingly dominant. For example, its probability accounts for about 87.6\% at 58 K. At lower temperature, its probability becomes even larger. As for the $|N = 0, J = 0.0\rangle$ and $|N = 1, J = 3.5\rangle$ atomic eigenstates, the probabilities are approximately 1.5\% and 0.7\% at 58 K, respectively. The probabilities for the other atomic eigenstates are quite small. Surprisingly, the evolutions of these atomic eigenstate probabilities with respect to temperature are not monotonic. Specially, the probabilities for the $|N = 1, J = 2.5\rangle$ and $|N = 0, J = 0.0\rangle$ atomic eigenstates drop quickly for $T < 20$ K and then grow up slowly for $T > 20$ K, whereas the one for the $|N = 1, J = 3.5\rangle$ atomic eigenstate shows a broad ``hump" in this region and reaches its maximum value near $T = 20$ K.

If we try to sum up the probabilities of atomic eigenstates with different $N$, we can obtain the distribution of 4$f$ electronic configurations, which will provide some useful information about the 4$f$ valence state fluctuations of the system. The change in atomic eigenstates probabilities must manifest itself in the electronic configurations and thus the valence state fluctuations. Fig.~\ref{fig:prob1}(b) shows the distribution of 4$f$ electronic configurations against temperature. Apparently, the $4f^{1}$ electronic configuration is ruling ($\sim 90\%$) all the time. The $4f^{2}$ and $4f^{0}$ electronic configurations are less important ($< 10\%$ and $< 2\%$, respectively). The $4f^{3}$ electronic configuration is rarely visited and can be ignored. Such a distribution is archetypal for Ce-based heavy fermion compounds~\cite{PhysRevB.81.195107,PhysRevB.94.075132} and well consistent with what we have learnt from the atomic eigenstates probabilities. Analogously, we also observe some anomalies near $T = 20$ K. In this region, the weight for the $4f^{2}$ electronic configuration reaches its maximum value, while on the contrary the one for the $4f^{1}$ electronic configuration is in its minimum value. In other words, the 4$f$ valence state fluctuation becomes strongest near $T = 20$ K~\cite{valence}. We believe that the sudden changes seen in the atomic eigenstate probabilities, distributions of electronic configurations, and valence state fluctuations could be explained as fingerprints or precursors of some kind of unknown transitions or crossovers, and have a tight relationship with the unusual features exhibited in the momentum-resolved spectral functions and density of states of CeB$_{6}$ (see Fig.~\ref{fig:akw}). 
 
\emph{$T$-dependent physical observables.} Now let us look at the temperature dependence of the other physical observables. First, we concentrate on the averaged 4$f$ occupancy $\langle N_{f}\rangle$, and the averaged total angular momentum $\langle J \rangle$. They were calculated using the following equations:
\begin{equation}
\langle N_{f}\rangle = \sum_{\Gamma} p_{\Gamma} N_{\Gamma},
\end{equation}
and
\begin{equation}
\langle J \rangle = \sum_{\Gamma} p_{\Gamma} J_{\Gamma}.
\end{equation}
Here $\Gamma$ denotes the index of atomic eigenstate, $p_{\Gamma}$ means the probability for the atomic eigenstate $|\Gamma\rangle$, $N_{\Gamma}$ and $J_{\Gamma}$ are the 4$f$ occupancy and total angular momentum for $|\Gamma \rangle$, respectively. As is clearly seen in Fig.~\ref{fig:prob2}(a), $\langle N_{f}\rangle$ and $\langle J \rangle$ firstly increase with respect to the temperature until a critical point, and then decline rapidly, showing a ``peak" in the vicinity of $T = 20$ K, which are in accordance with the previous discussion about the redistribution of the atomic eigenstate probabilities.

The total energy (and the corresponding free energy and entropy) should be temperature-dependent as well. How to compute them precisely is a big challenge for the DFT + DMFT calculations. Next, we try to evaluate the DFT + DMFT total energy $E_{\text{DFT + DMFT}}$ using the approach proposed by Haule \emph{et al.} very recently~\cite{PhysRevB.81.195107,PhysRevLett.115.256402}. The calculated results are shown in Fig.~\ref{fig:prob2}(b). We find that the $E_{\text{DFT + DMFT}}$ shows a deep ``valley" near 20 K. Noticed that the kinetic and potential energies of the system exhibit similar changes near $T = 20$ K as well (not shown here). 

\section{discussion\label{sec:disc}}

\begin{figure}[t]
\centering
\includegraphics[width=\columnwidth]{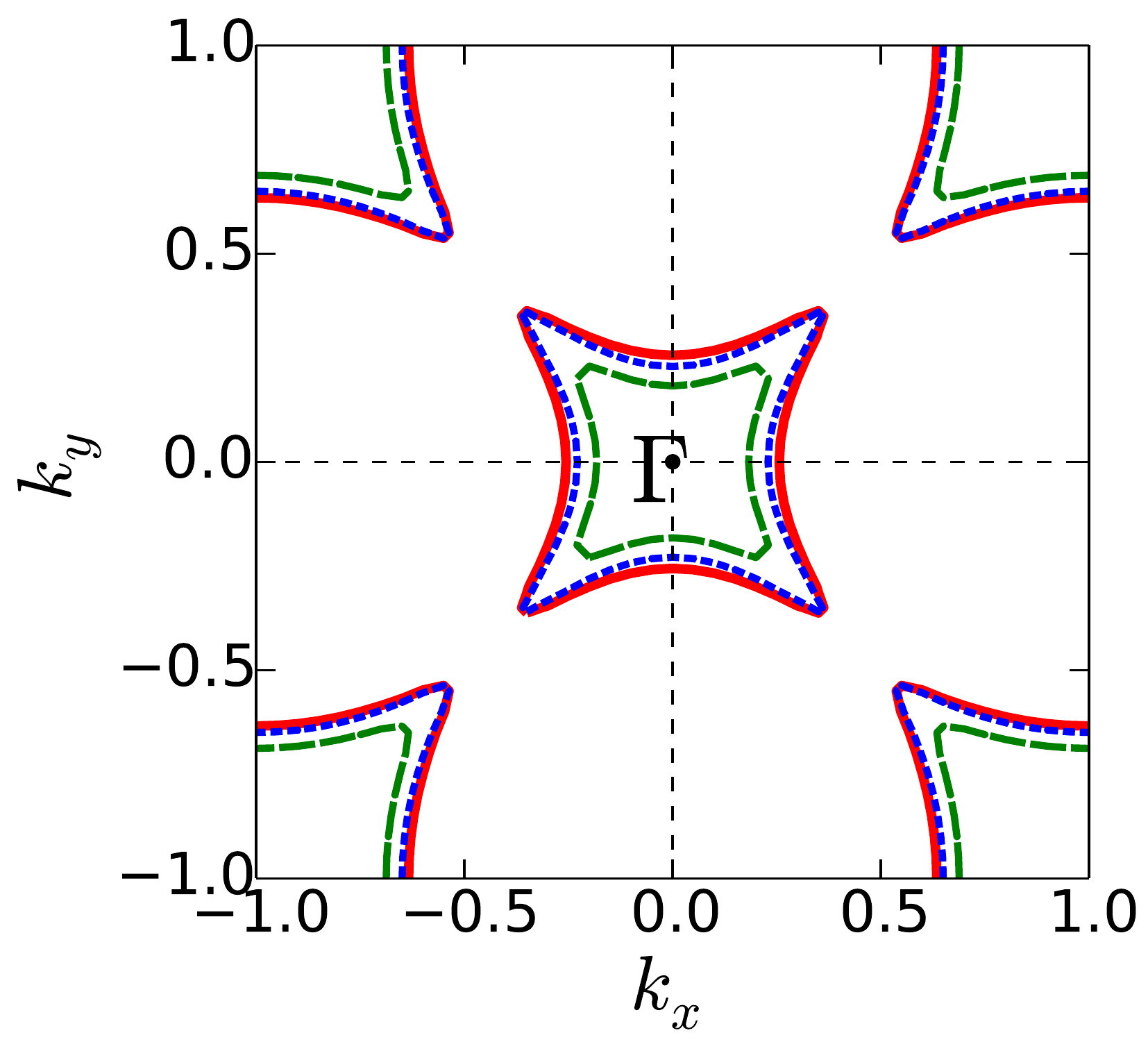}
\caption{(Color online). The temperature-dependent 2D Fermi surfaces (at the $k_x - k_y$ plane, i.e., $k_z = 0$). Calculated by the DFT + DMFT method under ambient pressure. The solid red, long-dashed green, and short-dashed blue lines denote the cases for $T = 38.7$, 23.2, and 11.6 K, respectively. Clearly, when $T \sim 20$ K, the area enclosed by the Fermi surface is the smallest. \label{fig:fs}}      
\end{figure}

In summary, all of the physical observables for CeB$_{6}$ studied in the present work (including the band structure, density of states, valence state histogram, electronic configuration, $\langle N_f \rangle$, $\langle J \rangle$, and $E_{\text{DFT + DMFT}}$, and so on) show unexpected changes or singularities near $T = 20$ K. If these results are correct instead of numerical fluctuations (actually, the bulk properties and electronic structures at $T \sim 17$ K have been successfully reproduced by us using the DFT + DMFT method, see main texts in Sec.~\ref{subsec:bench}), this makes sense and probably implies some kind of transitions or crossovers.     

Next we presume that the transition at $T \sim 20$ K really exists. So what is the possible mechanism for it? First of all, the magnetic structure of CeB$_{6}$ is extremely complicated~\cite{EFFANTIN1985145,KAWAKAMI1980435,JPSJ.52.728,PhysRevB.69.054415}, and there is complex competition among different multipolar orders/interactions (dipolar, quadrupolar, and octupolar) at low temperature~\cite{ERKELENS198761,Kobayashi281,PhysRevLett.103.017203}. Initially, it seems that the transition near 20 K is magnetic. However, the lowest temperature accessed in the present calculations ($\sim 10$ K) was much larger than either of $T_{\text{K}}$, $T_{\text{Q}}$, and $T_{\text{N}}$. Only the paramagnetic solutions were retained and no magnetic instability was observed in the calculations. So it is impossible to be a magnetic phase transition. Second, in some Ce-based heavy fermion systems, a 4$f$ localized-itinerant transition may take place if the external pressure or temperature reaches its critical point~\cite{shim:1615,PhysRevB.94.075132}. As is seen in Fig.~\ref{fig:akw}(g), there is no distinct and well-separated Kondo resonance peak in the Fermi level and the corresponding spectral intensity at $\omega = 0$ remains weak, which means that 4$f$ electrons in CeB$_{6}$ keep localized. As $T \sim 20$ K, small peak appears in the Fermi level. But it is difficult to establish a firm relationship between it and the localized-itinerant transition. Third, some materials would exhibit pressure- or temperature-driven electronic Lifshitz transition. In order to exclude or accept this possibility, we try to calculate the temperature-dependent Fermi surface of CeB$_{6}$. In Fig.~\ref{fig:fs}, the calculated Fermi surface on the $k_z = 0$ plane is shown. There is only one band crossing the Fermi level, so that the resulting Fermi surface is very simple. Obviously, though the volume or size of the Fermi surface is changed slightly with respect to temperature, its topology isn't. As a consequence, an electronic topological transition or electronic Lifshitz transition is excluded, at least at the temperature range we studied. Fourth, no superconducting behavior and formation of spin (or charge) density wave for $T > 10$ K have been reported in the literatures until now. So these probabilities are ruled out~\cite{yuan:2016}. Finally, we performed DFT + DMFT calculations for CeB$_{6}$ under ambient pressure. The obtained Matsubara self-energy functions suggest that the system stays in a non-Fermi-liquid state all the time. So that the likelihood for a temperature-driven non-Fermi-liquid to Fermi-liquid transition is eliminated. 

In a word, the temperature-dependent and abnormal behaviors exhibited in the physical observables of CeB$_{6}$ may indicate the entrance to a ``hidden" ordered phase, which can not be classified into any known phase of CeB$_6$. We don't have a rational explanation for this ``hidden'' phase transition yet. In order to unveil the secrets behind this phase transition, further experimental and theoretical studies are highly demanded.

\section{concluding remarks\label{sec:summary}}
In the present paper, we employed the \emph{ab initio} many-body approach, namely, the charge fully self-consistent DFT + DMFT method, to explore the temperature-dependent electronic structures of heavy fermion compound CeB$_{6}$ thoroughly. The calculated bulk properties and band structures (at 16.6 K) agree quite well with the available experimental data and ARPES results. The momentum-resolved spectral functions, density of states, 4$f$ valence state fluctuation, 4$f$ occupancy $\langle N_{f} \rangle$, total angular momentum $\langle J \rangle$, and total energy are calculated in the temperature range of $10$ K $< T < 120$ K. We find that the evolution of the physical observables with respect to temperature shows clear discontinuity when $T$ approaches 20 K. It could be a signature of a ``hidden" phase transition which has not been identified previously by either experiments or theoretical calculations. Thus, our calculated results can be considered as critical predictions and deserve further examinations. 

Finally, we would like to note that the ``hidden order'' and ``hidden ordered phase'' problems in the 5$f$ heavy fermion system URu$_{2}$Si$_{2}$ are very hot topics in the condense matter physics and have attracted many attentions in the last decades~\cite{rebecca:2014,RevModPhys.83.1301}. Many theories and methods have been developed for it. We think that the heavy fermion compound CeB$_{6}$ will provide another chance to gain insight on the hidden physics. 


\begin{acknowledgments}
We thank Prof. Yi-feng Yang for helpful discussions. This work was supported by the Natural Science Foundation of China (No.~11504340), Foundation of President of China Academy of Engineering Physics (No.~YZ2015012), Discipline Development Fund Project of Science and Technology on Surface Physics and Chemistry Laboratory (No.~201502), and Science Challenge Project of China (No.~JCKY2016212A56401).
\end{acknowledgments}


\bibliography{ceb6}

\begin{thebibliography}{56}%
\makeatletter
\providecommand \@ifxundefined [1]{%
 \@ifx{#1\undefined}
}%
\providecommand \@ifnum [1]{%
 \ifnum #1\expandafter \@firstoftwo
 \else \expandafter \@secondoftwo
 \fi
}%
\providecommand \@ifx [1]{%
 \ifx #1\expandafter \@firstoftwo
 \else \expandafter \@secondoftwo
 \fi
}%
\providecommand \natexlab [1]{#1}%
\providecommand \enquote  [1]{``#1''}%
\providecommand \bibnamefont  [1]{#1}%
\providecommand \bibfnamefont [1]{#1}%
\providecommand \citenamefont [1]{#1}%
\providecommand \href@noop [0]{\@secondoftwo}%
\providecommand \href [0]{\begingroup \@sanitize@url \@href}%
\providecommand \@href[1]{\@@startlink{#1}\@@href}%
\providecommand \@@href[1]{\endgroup#1\@@endlink}%
\providecommand \@sanitize@url [0]{\catcode `\\12\catcode `\$12\catcode
  `\&12\catcode `\#12\catcode `\^12\catcode `\_12\catcode `\%12\relax}%
\providecommand \@@startlink[1]{}%
\providecommand \@@endlink[0]{}%
\providecommand \url  [0]{\begingroup\@sanitize@url \@url }%
\providecommand \@url [1]{\endgroup\@href {#1}{\urlprefix }}%
\providecommand \urlprefix  [0]{URL }%
\providecommand \Eprint [0]{\href }%
\providecommand \doibase [0]{http://dx.doi.org/}%
\providecommand \selectlanguage [0]{\@gobble}%
\providecommand \bibinfo  [0]{\@secondoftwo}%
\providecommand \bibfield  [0]{\@secondoftwo}%
\providecommand \translation [1]{[#1]}%
\providecommand \BibitemOpen [0]{}%
\providecommand \bibitemStop [0]{}%
\providecommand \bibitemNoStop [0]{.\EOS\space}%
\providecommand \EOS [0]{\spacefactor3000\relax}%
\providecommand \BibitemShut  [1]{\csname bibitem#1\endcsname}%
\let\auto@bib@innerbib\@empty
\bibitem [{\citenamefont {Dordevic}\ \emph {et~al.}(2001)\citenamefont
  {Dordevic}, \citenamefont {Basov}, \citenamefont {Dilley}, \citenamefont
  {Bauer},\ and\ \citenamefont {Maple}}]{PhysRevLett.86.684}%
  \BibitemOpen
  \bibfield  {author} {\bibinfo {author} {\bibfnamefont {S.~V.}\ \bibnamefont
  {Dordevic}}, \bibinfo {author} {\bibfnamefont {D.~N.}\ \bibnamefont {Basov}},
  \bibinfo {author} {\bibfnamefont {N.~R.}\ \bibnamefont {Dilley}}, \bibinfo
  {author} {\bibfnamefont {E.~D.}\ \bibnamefont {Bauer}}, \ and\ \bibinfo
  {author} {\bibfnamefont {M.~B.}\ \bibnamefont {Maple}},\ }\href {\doibase
  10.1103/PhysRevLett.86.684} {\bibfield  {journal} {\bibinfo  {journal} {Phys.
  Rev. Lett.}\ }\textbf {\bibinfo {volume} {86}},\ \bibinfo {pages} {684}
  (\bibinfo {year} {2001})}\BibitemShut {NoStop}%
\bibitem [{\citenamefont {Im}\ \emph {et~al.}(2008)\citenamefont {Im},
  \citenamefont {Ito}, \citenamefont {Kim}, \citenamefont {Kimura},
  \citenamefont {Lee}, \citenamefont {Hong}, \citenamefont {Kwon},
  \citenamefont {Yasui},\ and\ \citenamefont
  {Yamagami}}]{PhysRevLett.100.176402}%
  \BibitemOpen
  \bibfield  {author} {\bibinfo {author} {\bibfnamefont {H.~J.}\ \bibnamefont
  {Im}}, \bibinfo {author} {\bibfnamefont {T.}~\bibnamefont {Ito}}, \bibinfo
  {author} {\bibfnamefont {H.-D.}\ \bibnamefont {Kim}}, \bibinfo {author}
  {\bibfnamefont {S.}~\bibnamefont {Kimura}}, \bibinfo {author} {\bibfnamefont
  {K.~E.}\ \bibnamefont {Lee}}, \bibinfo {author} {\bibfnamefont {J.~B.}\
  \bibnamefont {Hong}}, \bibinfo {author} {\bibfnamefont {Y.~S.}\ \bibnamefont
  {Kwon}}, \bibinfo {author} {\bibfnamefont {A.}~\bibnamefont {Yasui}}, \ and\
  \bibinfo {author} {\bibfnamefont {H.}~\bibnamefont {Yamagami}},\ }\href
  {\doibase 10.1103/PhysRevLett.100.176402} {\bibfield  {journal} {\bibinfo
  {journal} {Phys. Rev. Lett.}\ }\textbf {\bibinfo {volume} {100}},\ \bibinfo
  {pages} {176402} (\bibinfo {year} {2008})}\BibitemShut {NoStop}%
\bibitem [{\citenamefont {Stewart}(1984)}]{RevModPhys.56.755}%
  \BibitemOpen
  \bibfield  {author} {\bibinfo {author} {\bibfnamefont {G.~R.}\ \bibnamefont
  {Stewart}},\ }\href {\doibase 10.1103/RevModPhys.56.755} {\bibfield
  {journal} {\bibinfo  {journal} {Rev. Mod. Phys.}\ }\textbf {\bibinfo {volume}
  {56}},\ \bibinfo {pages} {755} (\bibinfo {year} {1984})}\BibitemShut
  {NoStop}%
\bibitem [{\citenamefont {Sigrist}\ and\ \citenamefont
  {Ueda}(1991)}]{RevModPhys.63.239}%
  \BibitemOpen
  \bibfield  {author} {\bibinfo {author} {\bibfnamefont {M.}~\bibnamefont
  {Sigrist}}\ and\ \bibinfo {author} {\bibfnamefont {K.}~\bibnamefont {Ueda}},\
  }\href {\doibase 10.1103/RevModPhys.63.239} {\bibfield  {journal} {\bibinfo
  {journal} {Rev. Mod. Phys.}\ }\textbf {\bibinfo {volume} {63}},\ \bibinfo
  {pages} {239} (\bibinfo {year} {1991})}\BibitemShut {NoStop}%
\bibitem [{\citenamefont {Bauer}\ \emph {et~al.}(2004)\citenamefont {Bauer},
  \citenamefont {Hilscher}, \citenamefont {Michor}, \citenamefont {Paul},
  \citenamefont {Scheidt}, \citenamefont {Gribanov}, \citenamefont {Seropegin},
  \citenamefont {No\"el}, \citenamefont {Sigrist},\ and\ \citenamefont
  {Rogl}}]{PhysRevLett.92.027003}%
  \BibitemOpen
  \bibfield  {author} {\bibinfo {author} {\bibfnamefont {E.}~\bibnamefont
  {Bauer}}, \bibinfo {author} {\bibfnamefont {G.}~\bibnamefont {Hilscher}},
  \bibinfo {author} {\bibfnamefont {H.}~\bibnamefont {Michor}}, \bibinfo
  {author} {\bibfnamefont {C.}~\bibnamefont {Paul}}, \bibinfo {author}
  {\bibfnamefont {E.~W.}\ \bibnamefont {Scheidt}}, \bibinfo {author}
  {\bibfnamefont {A.}~\bibnamefont {Gribanov}}, \bibinfo {author}
  {\bibfnamefont {Y.}~\bibnamefont {Seropegin}}, \bibinfo {author}
  {\bibfnamefont {H.}~\bibnamefont {No\"el}}, \bibinfo {author} {\bibfnamefont
  {M.}~\bibnamefont {Sigrist}}, \ and\ \bibinfo {author} {\bibfnamefont
  {P.}~\bibnamefont {Rogl}},\ }\href {\doibase 10.1103/PhysRevLett.92.027003}
  {\bibfield  {journal} {\bibinfo  {journal} {Phys. Rev. Lett.}\ }\textbf
  {\bibinfo {volume} {92}},\ \bibinfo {pages} {027003} (\bibinfo {year}
  {2004})}\BibitemShut {NoStop}%
\bibitem [{\citenamefont {Izawa}\ \emph {et~al.}(2001)\citenamefont {Izawa},
  \citenamefont {Yamaguchi}, \citenamefont {Matsuda}, \citenamefont {Shishido},
  \citenamefont {Settai},\ and\ \citenamefont {Onuki}}]{PhysRevLett.87.057002}%
  \BibitemOpen
  \bibfield  {author} {\bibinfo {author} {\bibfnamefont {K.}~\bibnamefont
  {Izawa}}, \bibinfo {author} {\bibfnamefont {H.}~\bibnamefont {Yamaguchi}},
  \bibinfo {author} {\bibfnamefont {Y.}~\bibnamefont {Matsuda}}, \bibinfo
  {author} {\bibfnamefont {H.}~\bibnamefont {Shishido}}, \bibinfo {author}
  {\bibfnamefont {R.}~\bibnamefont {Settai}}, \ and\ \bibinfo {author}
  {\bibfnamefont {Y.}~\bibnamefont {Onuki}},\ }\href {\doibase
  10.1103/PhysRevLett.87.057002} {\bibfield  {journal} {\bibinfo  {journal}
  {Phys. Rev. Lett.}\ }\textbf {\bibinfo {volume} {87}},\ \bibinfo {pages}
  {057002} (\bibinfo {year} {2001})}\BibitemShut {NoStop}%
\bibitem [{\citenamefont {Paglione}\ \emph {et~al.}(2016)\citenamefont
  {Paglione}, \citenamefont {Tanatar}, \citenamefont {Reid}, \citenamefont
  {Shakeripour}, \citenamefont {Petrovic},\ and\ \citenamefont
  {Taillefer}}]{PhysRevLett.117.016601}%
  \BibitemOpen
  \bibfield  {author} {\bibinfo {author} {\bibfnamefont {J.}~\bibnamefont
  {Paglione}}, \bibinfo {author} {\bibfnamefont {M.~A.}\ \bibnamefont
  {Tanatar}}, \bibinfo {author} {\bibfnamefont {J.-P.}\ \bibnamefont {Reid}},
  \bibinfo {author} {\bibfnamefont {H.}~\bibnamefont {Shakeripour}}, \bibinfo
  {author} {\bibfnamefont {C.}~\bibnamefont {Petrovic}}, \ and\ \bibinfo
  {author} {\bibfnamefont {L.}~\bibnamefont {Taillefer}},\ }\href {\doibase
  10.1103/PhysRevLett.117.016601} {\bibfield  {journal} {\bibinfo  {journal}
  {Phys. Rev. Lett.}\ }\textbf {\bibinfo {volume} {117}},\ \bibinfo {pages}
  {016601} (\bibinfo {year} {2016})}\BibitemShut {NoStop}%
\bibitem [{\citenamefont {Ishii}\ \emph {et~al.}(2016)\citenamefont {Ishii},
  \citenamefont {Toda}, \citenamefont {Hanaoka}, \citenamefont {Tokiwa},
  \citenamefont {Shimozawa}, \citenamefont {Kasahara}, \citenamefont {Endo},
  \citenamefont {Terashima}, \citenamefont {Nevidomskyy}, \citenamefont
  {Shibauchi},\ and\ \citenamefont {Matsuda}}]{PhysRevLett.116.206401}%
  \BibitemOpen
  \bibfield  {author} {\bibinfo {author} {\bibfnamefont {T.}~\bibnamefont
  {Ishii}}, \bibinfo {author} {\bibfnamefont {R.}~\bibnamefont {Toda}},
  \bibinfo {author} {\bibfnamefont {Y.}~\bibnamefont {Hanaoka}}, \bibinfo
  {author} {\bibfnamefont {Y.}~\bibnamefont {Tokiwa}}, \bibinfo {author}
  {\bibfnamefont {M.}~\bibnamefont {Shimozawa}}, \bibinfo {author}
  {\bibfnamefont {Y.}~\bibnamefont {Kasahara}}, \bibinfo {author}
  {\bibfnamefont {R.}~\bibnamefont {Endo}}, \bibinfo {author} {\bibfnamefont
  {T.}~\bibnamefont {Terashima}}, \bibinfo {author} {\bibfnamefont {A.~H.}\
  \bibnamefont {Nevidomskyy}}, \bibinfo {author} {\bibfnamefont
  {T.}~\bibnamefont {Shibauchi}}, \ and\ \bibinfo {author} {\bibfnamefont
  {Y.}~\bibnamefont {Matsuda}},\ }\href {\doibase
  10.1103/PhysRevLett.116.206401} {\bibfield  {journal} {\bibinfo  {journal}
  {Phys. Rev. Lett.}\ }\textbf {\bibinfo {volume} {116}},\ \bibinfo {pages}
  {206401} (\bibinfo {year} {2016})}\BibitemShut {NoStop}%
\bibitem [{\citenamefont {Lin}\ \emph {et~al.}(2015)\citenamefont {Lin},
  \citenamefont {Shirer}, \citenamefont {Crocker}, \citenamefont {Dioguardi},
  \citenamefont {Lawson}, \citenamefont {Bush}, \citenamefont {Klavins},\ and\
  \citenamefont {Curro}}]{PhysRevB.92.155147}%
  \BibitemOpen
  \bibfield  {author} {\bibinfo {author} {\bibfnamefont {C.~H.}\ \bibnamefont
  {Lin}}, \bibinfo {author} {\bibfnamefont {K.~R.}\ \bibnamefont {Shirer}},
  \bibinfo {author} {\bibfnamefont {J.}~\bibnamefont {Crocker}}, \bibinfo
  {author} {\bibfnamefont {A.~P.}\ \bibnamefont {Dioguardi}}, \bibinfo {author}
  {\bibfnamefont {M.~M.}\ \bibnamefont {Lawson}}, \bibinfo {author}
  {\bibfnamefont {B.~T.}\ \bibnamefont {Bush}}, \bibinfo {author}
  {\bibfnamefont {P.}~\bibnamefont {Klavins}}, \ and\ \bibinfo {author}
  {\bibfnamefont {N.~J.}\ \bibnamefont {Curro}},\ }\href {\doibase
  10.1103/PhysRevB.92.155147} {\bibfield  {journal} {\bibinfo  {journal} {Phys.
  Rev. B}\ }\textbf {\bibinfo {volume} {92}},\ \bibinfo {pages} {155147}
  (\bibinfo {year} {2015})}\BibitemShut {NoStop}%
\bibitem [{\citenamefont {Watanabe}\ and\ \citenamefont
  {Ogata}(2007)}]{PhysRevLett.99.136401}%
  \BibitemOpen
  \bibfield  {author} {\bibinfo {author} {\bibfnamefont {H.}~\bibnamefont
  {Watanabe}}\ and\ \bibinfo {author} {\bibfnamefont {M.}~\bibnamefont
  {Ogata}},\ }\href {\doibase 10.1103/PhysRevLett.99.136401} {\bibfield
  {journal} {\bibinfo  {journal} {Phys. Rev. Lett.}\ }\textbf {\bibinfo
  {volume} {99}},\ \bibinfo {pages} {136401} (\bibinfo {year}
  {2007})}\BibitemShut {NoStop}%
\bibitem [{\citenamefont {Danzenb\"acher}\ \emph {et~al.}(2009)\citenamefont
  {Danzenb\"acher}, \citenamefont {Vyalikh}, \citenamefont {Kucherenko},
  \citenamefont {Kade}, \citenamefont {Laubschat}, \citenamefont
  {Caroca-Canales}, \citenamefont {Krellner}, \citenamefont {Geibel},
  \citenamefont {Fedorov}, \citenamefont {Dessau}, \citenamefont {Follath},
  \citenamefont {Eberhardt},\ and\ \citenamefont
  {Molodtsov}}]{PhysRevLett.102.026403}%
  \BibitemOpen
  \bibfield  {author} {\bibinfo {author} {\bibfnamefont {S.}~\bibnamefont
  {Danzenb\"acher}}, \bibinfo {author} {\bibfnamefont {D.~V.}\ \bibnamefont
  {Vyalikh}}, \bibinfo {author} {\bibfnamefont {Y.}~\bibnamefont {Kucherenko}},
  \bibinfo {author} {\bibfnamefont {A.}~\bibnamefont {Kade}}, \bibinfo {author}
  {\bibfnamefont {C.}~\bibnamefont {Laubschat}}, \bibinfo {author}
  {\bibfnamefont {N.}~\bibnamefont {Caroca-Canales}}, \bibinfo {author}
  {\bibfnamefont {C.}~\bibnamefont {Krellner}}, \bibinfo {author}
  {\bibfnamefont {C.}~\bibnamefont {Geibel}}, \bibinfo {author} {\bibfnamefont
  {A.~V.}\ \bibnamefont {Fedorov}}, \bibinfo {author} {\bibfnamefont {D.~S.}\
  \bibnamefont {Dessau}}, \bibinfo {author} {\bibfnamefont {R.}~\bibnamefont
  {Follath}}, \bibinfo {author} {\bibfnamefont {W.}~\bibnamefont {Eberhardt}},
  \ and\ \bibinfo {author} {\bibfnamefont {S.~L.}\ \bibnamefont {Molodtsov}},\
  }\href {\doibase 10.1103/PhysRevLett.102.026403} {\bibfield  {journal}
  {\bibinfo  {journal} {Phys. Rev. Lett.}\ }\textbf {\bibinfo {volume} {102}},\
  \bibinfo {pages} {026403} (\bibinfo {year} {2009})}\BibitemShut {NoStop}%
\bibitem [{\citenamefont {Lee}\ \emph {et~al.}(1987)\citenamefont {Lee},
  \citenamefont {Ku},\ and\ \citenamefont {Shelton}}]{PhysRevB.36.5739}%
  \BibitemOpen
  \bibfield  {author} {\bibinfo {author} {\bibfnamefont {W.~H.}\ \bibnamefont
  {Lee}}, \bibinfo {author} {\bibfnamefont {H.~C.}\ \bibnamefont {Ku}}, \ and\
  \bibinfo {author} {\bibfnamefont {R.~N.}\ \bibnamefont {Shelton}},\ }\href
  {\doibase 10.1103/PhysRevB.36.5739} {\bibfield  {journal} {\bibinfo
  {journal} {Phys. Rev. B}\ }\textbf {\bibinfo {volume} {36}},\ \bibinfo
  {pages} {5739} (\bibinfo {year} {1987})}\BibitemShut {NoStop}%
\bibitem [{\citenamefont {Weng}\ \emph {et~al.}(2016)\citenamefont {Weng},
  \citenamefont {Smidman}, \citenamefont {Jiao}, \citenamefont {Lu},\ and\
  \citenamefont {Yuan}}]{yuan:2016}%
  \BibitemOpen
  \bibfield  {author} {\bibinfo {author} {\bibfnamefont {Z.~F.}\ \bibnamefont
  {Weng}}, \bibinfo {author} {\bibfnamefont {M.}~\bibnamefont {Smidman}},
  \bibinfo {author} {\bibfnamefont {L.}~\bibnamefont {Jiao}}, \bibinfo {author}
  {\bibfnamefont {X.}~\bibnamefont {Lu}}, \ and\ \bibinfo {author}
  {\bibfnamefont {H.~Q.}\ \bibnamefont {Yuan}},\ }\href
  {http://stacks.iop.org/0034-4885/79/i=9/a=094503} {\bibfield  {journal}
  {\bibinfo  {journal} {Rep. Prog. Phys.}\ }\textbf {\bibinfo {volume} {79}},\
  \bibinfo {pages} {094503} (\bibinfo {year} {2016})}\BibitemShut {NoStop}%
\bibitem [{\citenamefont {Takabatake}\ \emph {et~al.}(1990)\citenamefont
  {Takabatake}, \citenamefont {Teshima}, \citenamefont {Fujii}, \citenamefont
  {Nishigori}, \citenamefont {Suzuki}, \citenamefont {Fujita}, \citenamefont
  {Yamaguchi}, \citenamefont {Sakurai},\ and\ \citenamefont
  {Jaccard}}]{PhysRevB.41.9607}%
  \BibitemOpen
  \bibfield  {author} {\bibinfo {author} {\bibfnamefont {T.}~\bibnamefont
  {Takabatake}}, \bibinfo {author} {\bibfnamefont {F.}~\bibnamefont {Teshima}},
  \bibinfo {author} {\bibfnamefont {H.}~\bibnamefont {Fujii}}, \bibinfo
  {author} {\bibfnamefont {S.}~\bibnamefont {Nishigori}}, \bibinfo {author}
  {\bibfnamefont {T.}~\bibnamefont {Suzuki}}, \bibinfo {author} {\bibfnamefont
  {T.}~\bibnamefont {Fujita}}, \bibinfo {author} {\bibfnamefont
  {Y.}~\bibnamefont {Yamaguchi}}, \bibinfo {author} {\bibfnamefont
  {J.}~\bibnamefont {Sakurai}}, \ and\ \bibinfo {author} {\bibfnamefont
  {D.}~\bibnamefont {Jaccard}},\ }\href {\doibase 10.1103/PhysRevB.41.9607}
  {\bibfield  {journal} {\bibinfo  {journal} {Phys. Rev. B}\ }\textbf {\bibinfo
  {volume} {41}},\ \bibinfo {pages} {9607} (\bibinfo {year}
  {1990})}\BibitemShut {NoStop}%
\bibitem [{\citenamefont {Choi}\ \emph {et~al.}(2013)\citenamefont {Choi},
  \citenamefont {Haule}, \citenamefont {Kotliar}, \citenamefont {Min},\ and\
  \citenamefont {Shim}}]{PhysRevB.88.125111}%
  \BibitemOpen
  \bibfield  {author} {\bibinfo {author} {\bibfnamefont {H.~C.}\ \bibnamefont
  {Choi}}, \bibinfo {author} {\bibfnamefont {K.}~\bibnamefont {Haule}},
  \bibinfo {author} {\bibfnamefont {G.}~\bibnamefont {Kotliar}}, \bibinfo
  {author} {\bibfnamefont {B.~I.}\ \bibnamefont {Min}}, \ and\ \bibinfo
  {author} {\bibfnamefont {J.~H.}\ \bibnamefont {Shim}},\ }\href {\doibase
  10.1103/PhysRevB.88.125111} {\bibfield  {journal} {\bibinfo  {journal} {Phys.
  Rev. B}\ }\textbf {\bibinfo {volume} {88}},\ \bibinfo {pages} {125111}
  (\bibinfo {year} {2013})}\BibitemShut {NoStop}%
\bibitem [{\citenamefont {Choi}\ \emph {et~al.}(2012)\citenamefont {Choi},
  \citenamefont {Min}, \citenamefont {Shim}, \citenamefont {Haule},\ and\
  \citenamefont {Kotliar}}]{PhysRevLett.108.016402}%
  \BibitemOpen
  \bibfield  {author} {\bibinfo {author} {\bibfnamefont {H.~C.}\ \bibnamefont
  {Choi}}, \bibinfo {author} {\bibfnamefont {B.~I.}\ \bibnamefont {Min}},
  \bibinfo {author} {\bibfnamefont {J.~H.}\ \bibnamefont {Shim}}, \bibinfo
  {author} {\bibfnamefont {K.}~\bibnamefont {Haule}}, \ and\ \bibinfo {author}
  {\bibfnamefont {G.}~\bibnamefont {Kotliar}},\ }\href {\doibase
  10.1103/PhysRevLett.108.016402} {\bibfield  {journal} {\bibinfo  {journal}
  {Phys. Rev. Lett.}\ }\textbf {\bibinfo {volume} {108}},\ \bibinfo {pages}
  {016402} (\bibinfo {year} {2012})}\BibitemShut {NoStop}%
\bibitem [{\citenamefont {Nomoto}\ and\ \citenamefont
  {Ikeda}(2014)}]{PhysRevB.90.125147}%
  \BibitemOpen
  \bibfield  {author} {\bibinfo {author} {\bibfnamefont {T.}~\bibnamefont
  {Nomoto}}\ and\ \bibinfo {author} {\bibfnamefont {H.}~\bibnamefont {Ikeda}},\
  }\href {\doibase 10.1103/PhysRevB.90.125147} {\bibfield  {journal} {\bibinfo
  {journal} {Phys. Rev. B}\ }\textbf {\bibinfo {volume} {90}},\ \bibinfo
  {pages} {125147} (\bibinfo {year} {2014})}\BibitemShut {NoStop}%
\bibitem [{\citenamefont {Blomberg}\ \emph {et~al.}(1991)\citenamefont
  {Blomberg}, \citenamefont {Merisalo}, \citenamefont {Korsukova},\ and\
  \citenamefont {Gurin}}]{BLOMBERG1991313}%
  \BibitemOpen
  \bibfield  {author} {\bibinfo {author} {\bibfnamefont {M.}~\bibnamefont
  {Blomberg}}, \bibinfo {author} {\bibfnamefont {M.}~\bibnamefont {Merisalo}},
  \bibinfo {author} {\bibfnamefont {M.}~\bibnamefont {Korsukova}}, \ and\
  \bibinfo {author} {\bibfnamefont {V.}~\bibnamefont {Gurin}},\ }\href
  {\doibase http://dx.doi.org/10.1016/0022-5088(91)90313-S} {\bibfield
  {journal} {\bibinfo  {journal} {J. Less-Common Metals}\ }\textbf {\bibinfo
  {volume} {168}},\ \bibinfo {pages} {313} (\bibinfo {year}
  {1991})}\BibitemShut {NoStop}%
\bibitem [{\citenamefont {Biasini}\ \emph {et~al.}(1994)\citenamefont
  {Biasini}, \citenamefont {Alam}, \citenamefont {Harima}, \citenamefont
  {Onuki}, \citenamefont {Fretwell},\ and\ \citenamefont {West}}]{jpcm7823}%
  \BibitemOpen
  \bibfield  {author} {\bibinfo {author} {\bibfnamefont {M.}~\bibnamefont
  {Biasini}}, \bibinfo {author} {\bibfnamefont {M.~A.}\ \bibnamefont {Alam}},
  \bibinfo {author} {\bibfnamefont {H.}~\bibnamefont {Harima}}, \bibinfo
  {author} {\bibfnamefont {Y.}~\bibnamefont {Onuki}}, \bibinfo {author}
  {\bibfnamefont {H.~M.}\ \bibnamefont {Fretwell}}, \ and\ \bibinfo {author}
  {\bibfnamefont {R.~N.}\ \bibnamefont {West}},\ }\href
  {http://stacks.iop.org/0953-8984/6/i=38/a=019} {\bibfield  {journal}
  {\bibinfo  {journal} {J. Phys.: Condens. Matter}\ }\textbf {\bibinfo {volume}
  {6}},\ \bibinfo {pages} {7823} (\bibinfo {year} {1994})}\BibitemShut
  {NoStop}%
\bibitem [{\citenamefont {Effantin}\ \emph {et~al.}(1985)\citenamefont
  {Effantin}, \citenamefont {Rossat-Mignod}, \citenamefont {Burlet},
  \citenamefont {Bartholin}, \citenamefont {Kunii},\ and\ \citenamefont
  {Kasuya}}]{EFFANTIN1985145}%
  \BibitemOpen
  \bibfield  {author} {\bibinfo {author} {\bibfnamefont {J.}~\bibnamefont
  {Effantin}}, \bibinfo {author} {\bibfnamefont {J.}~\bibnamefont
  {Rossat-Mignod}}, \bibinfo {author} {\bibfnamefont {P.}~\bibnamefont
  {Burlet}}, \bibinfo {author} {\bibfnamefont {H.}~\bibnamefont {Bartholin}},
  \bibinfo {author} {\bibfnamefont {S.}~\bibnamefont {Kunii}}, \ and\ \bibinfo
  {author} {\bibfnamefont {T.}~\bibnamefont {Kasuya}},\ }\href {\doibase
  http://dx.doi.org/10.1016/0304-8853(85)90382-8} {\bibfield  {journal}
  {\bibinfo  {journal} {J. Magn. Magn. Mater.}\ }\textbf {\bibinfo {volume}
  {47}},\ \bibinfo {pages} {145} (\bibinfo {year} {1985})}\BibitemShut
  {NoStop}%
\bibitem [{\citenamefont {Kawakami}\ \emph {et~al.}(1980)\citenamefont
  {Kawakami}, \citenamefont {Kunii}, \citenamefont {Komatsubara},\ and\
  \citenamefont {Kasuya}}]{KAWAKAMI1980435}%
  \BibitemOpen
  \bibfield  {author} {\bibinfo {author} {\bibfnamefont {M.}~\bibnamefont
  {Kawakami}}, \bibinfo {author} {\bibfnamefont {S.}~\bibnamefont {Kunii}},
  \bibinfo {author} {\bibfnamefont {T.}~\bibnamefont {Komatsubara}}, \ and\
  \bibinfo {author} {\bibfnamefont {T.}~\bibnamefont {Kasuya}},\ }\href
  {\doibase http://dx.doi.org/10.1016/0038-1098(80)90928-X} {\bibfield
  {journal} {\bibinfo  {journal} {Solid State Commun.}\ }\textbf {\bibinfo
  {volume} {36}},\ \bibinfo {pages} {435 } (\bibinfo {year}
  {1980})}\BibitemShut {NoStop}%
\bibitem [{\citenamefont {Takigawa}\ \emph {et~al.}(1983)\citenamefont
  {Takigawa}, \citenamefont {Yasuoka}, \citenamefont {Tanaka},\ and\
  \citenamefont {Ishizawa}}]{JPSJ.52.728}%
  \BibitemOpen
  \bibfield  {author} {\bibinfo {author} {\bibfnamefont {M.}~\bibnamefont
  {Takigawa}}, \bibinfo {author} {\bibfnamefont {H.}~\bibnamefont {Yasuoka}},
  \bibinfo {author} {\bibfnamefont {T.}~\bibnamefont {Tanaka}}, \ and\ \bibinfo
  {author} {\bibfnamefont {Y.}~\bibnamefont {Ishizawa}},\ }\href {\doibase
  10.1143/JPSJ.52.728} {\bibfield  {journal} {\bibinfo  {journal} {J. Phys.
  Soc. Jpn.}\ }\textbf {\bibinfo {volume} {52}},\ \bibinfo {pages} {728}
  (\bibinfo {year} {1983})}\BibitemShut {NoStop}%
\bibitem [{\citenamefont {Goodrich}\ \emph {et~al.}(2004)\citenamefont
  {Goodrich}, \citenamefont {Young}, \citenamefont {Hall}, \citenamefont
  {Balicas}, \citenamefont {Fisk}, \citenamefont {Harrison}, \citenamefont
  {Betts}, \citenamefont {Migliori}, \citenamefont {Woodward},\ and\
  \citenamefont {Lynn}}]{PhysRevB.69.054415}%
  \BibitemOpen
  \bibfield  {author} {\bibinfo {author} {\bibfnamefont {R.~G.}\ \bibnamefont
  {Goodrich}}, \bibinfo {author} {\bibfnamefont {D.~P.}\ \bibnamefont {Young}},
  \bibinfo {author} {\bibfnamefont {D.}~\bibnamefont {Hall}}, \bibinfo {author}
  {\bibfnamefont {L.}~\bibnamefont {Balicas}}, \bibinfo {author} {\bibfnamefont
  {Z.}~\bibnamefont {Fisk}}, \bibinfo {author} {\bibfnamefont {N.}~\bibnamefont
  {Harrison}}, \bibinfo {author} {\bibfnamefont {J.}~\bibnamefont {Betts}},
  \bibinfo {author} {\bibfnamefont {A.}~\bibnamefont {Migliori}}, \bibinfo
  {author} {\bibfnamefont {F.~M.}\ \bibnamefont {Woodward}}, \ and\ \bibinfo
  {author} {\bibfnamefont {J.~W.}\ \bibnamefont {Lynn}},\ }\href {\doibase
  10.1103/PhysRevB.69.054415} {\bibfield  {journal} {\bibinfo  {journal} {Phys.
  Rev. B}\ }\textbf {\bibinfo {volume} {69}},\ \bibinfo {pages} {054415}
  (\bibinfo {year} {2004})}\BibitemShut {NoStop}%
\bibitem [{\citenamefont {Kobayashi}\ \emph {et~al.}(2000)\citenamefont
  {Kobayashi}, \citenamefont {Hashimoto}, \citenamefont {Eda}, \citenamefont
  {Shimizu}, \citenamefont {Amaya},\ and\ \citenamefont
  {Onuki}}]{Kobayashi281}%
  \BibitemOpen
  \bibfield  {author} {\bibinfo {author} {\bibfnamefont {T.~C.}\ \bibnamefont
  {Kobayashi}}, \bibinfo {author} {\bibfnamefont {K.}~\bibnamefont
  {Hashimoto}}, \bibinfo {author} {\bibfnamefont {S.}~\bibnamefont {Eda}},
  \bibinfo {author} {\bibfnamefont {K.}~\bibnamefont {Shimizu}}, \bibinfo
  {author} {\bibfnamefont {K.}~\bibnamefont {Amaya}}, \ and\ \bibinfo {author}
  {\bibfnamefont {Y.}~\bibnamefont {Onuki}},\ }\href {\doibase
  10.1016/S0921-4526(99)00947-3} {\bibfield  {journal} {\bibinfo  {journal}
  {Physica B: Phys. Conden. Matter}\ }\textbf {\bibinfo {volume} {281}},\
  \bibinfo {pages} {553 } (\bibinfo {year} {2000})}\BibitemShut {NoStop}%
\bibitem [{\citenamefont {Yamamizu}\ \emph {et~al.}(2004)\citenamefont
  {Yamamizu}, \citenamefont {Endo}, \citenamefont {Nakayama}, \citenamefont
  {Kimura}, \citenamefont {Aoki},\ and\ \citenamefont
  {Kunii}}]{PhysRevB.69.014423}%
  \BibitemOpen
  \bibfield  {author} {\bibinfo {author} {\bibfnamefont {T.}~\bibnamefont
  {Yamamizu}}, \bibinfo {author} {\bibfnamefont {M.}~\bibnamefont {Endo}},
  \bibinfo {author} {\bibfnamefont {M.}~\bibnamefont {Nakayama}}, \bibinfo
  {author} {\bibfnamefont {N.}~\bibnamefont {Kimura}}, \bibinfo {author}
  {\bibfnamefont {H.}~\bibnamefont {Aoki}}, \ and\ \bibinfo {author}
  {\bibfnamefont {S.}~\bibnamefont {Kunii}},\ }\href {\doibase
  10.1103/PhysRevB.69.014423} {\bibfield  {journal} {\bibinfo  {journal} {Phys.
  Rev. B}\ }\textbf {\bibinfo {volume} {69}},\ \bibinfo {pages} {014423}
  (\bibinfo {year} {2004})}\BibitemShut {NoStop}%
\bibitem [{\citenamefont {Sluchanko}\ \emph {et~al.}(2007)\citenamefont
  {Sluchanko}, \citenamefont {Bogach}, \citenamefont {Glushkov}, \citenamefont
  {Demishev}, \citenamefont {Ivanov}, \citenamefont {Ignatov}, \citenamefont
  {Kuznetsov}, \citenamefont {Samarin}, \citenamefont {Semeno},\ and\
  \citenamefont {Shitsevalova}}]{Sluchanko2007}%
  \BibitemOpen
  \bibfield  {author} {\bibinfo {author} {\bibfnamefont {N.~E.}\ \bibnamefont
  {Sluchanko}}, \bibinfo {author} {\bibfnamefont {A.~V.}\ \bibnamefont
  {Bogach}}, \bibinfo {author} {\bibfnamefont {V.~V.}\ \bibnamefont
  {Glushkov}}, \bibinfo {author} {\bibfnamefont {S.~V.}\ \bibnamefont
  {Demishev}}, \bibinfo {author} {\bibfnamefont {V.~Y.}\ \bibnamefont
  {Ivanov}}, \bibinfo {author} {\bibfnamefont {M.~I.}\ \bibnamefont {Ignatov}},
  \bibinfo {author} {\bibfnamefont {A.~V.}\ \bibnamefont {Kuznetsov}}, \bibinfo
  {author} {\bibfnamefont {N.~A.}\ \bibnamefont {Samarin}}, \bibinfo {author}
  {\bibfnamefont {A.~V.}\ \bibnamefont {Semeno}}, \ and\ \bibinfo {author}
  {\bibfnamefont {N.~Y.}\ \bibnamefont {Shitsevalova}},\ }\href {\doibase
  10.1134/S106377610701013X} {\bibfield  {journal} {\bibinfo  {journal} {J.
  Exp. Theor. Phys.}\ }\textbf {\bibinfo {volume} {104}},\ \bibinfo {pages}
  {120} (\bibinfo {year} {2007})}\BibitemShut {NoStop}%
\bibitem [{\citenamefont {Brandt}\ \emph {et~al.}(1985)\citenamefont {Brandt},
  \citenamefont {Moshchalkov}, \citenamefont {Pashkevich}, \citenamefont
  {Vybornov}, \citenamefont {Semenov}, \citenamefont {Kolobyanina},
  \citenamefont {Konovalova},\ and\ \citenamefont {Paderno}}]{BRANDT1985937}%
  \BibitemOpen
  \bibfield  {author} {\bibinfo {author} {\bibfnamefont {N.}~\bibnamefont
  {Brandt}}, \bibinfo {author} {\bibfnamefont {V.}~\bibnamefont {Moshchalkov}},
  \bibinfo {author} {\bibfnamefont {S.}~\bibnamefont {Pashkevich}}, \bibinfo
  {author} {\bibfnamefont {M.}~\bibnamefont {Vybornov}}, \bibinfo {author}
  {\bibfnamefont {M.}~\bibnamefont {Semenov}}, \bibinfo {author} {\bibfnamefont
  {T.}~\bibnamefont {Kolobyanina}}, \bibinfo {author} {\bibfnamefont
  {E.}~\bibnamefont {Konovalova}}, \ and\ \bibinfo {author} {\bibfnamefont
  {Y.}~\bibnamefont {Paderno}},\ }\href {\doibase
  http://dx.doi.org/10.1016/S0038-1098(85)80029-6} {\bibfield  {journal}
  {\bibinfo  {journal} {Solid State Commun.}\ }\textbf {\bibinfo {volume}
  {56}},\ \bibinfo {pages} {937 } (\bibinfo {year} {1985})}\BibitemShut
  {NoStop}%
\bibitem [{\citenamefont {Erkelens}\ \emph {et~al.}(1987)\citenamefont
  {Erkelens}, \citenamefont {Regnault}, \citenamefont {Burlet}, \citenamefont
  {Rossat-Mignod}, \citenamefont {Kunii},\ and\ \citenamefont
  {Kasuya}}]{ERKELENS198761}%
  \BibitemOpen
  \bibfield  {author} {\bibinfo {author} {\bibfnamefont {W.}~\bibnamefont
  {Erkelens}}, \bibinfo {author} {\bibfnamefont {L.}~\bibnamefont {Regnault}},
  \bibinfo {author} {\bibfnamefont {P.}~\bibnamefont {Burlet}}, \bibinfo
  {author} {\bibfnamefont {J.}~\bibnamefont {Rossat-Mignod}}, \bibinfo {author}
  {\bibfnamefont {S.}~\bibnamefont {Kunii}}, \ and\ \bibinfo {author}
  {\bibfnamefont {T.}~\bibnamefont {Kasuya}},\ }\href {\doibase
  http://dx.doi.org/10.1016/0304-8853(87)90522-1} {\bibfield  {journal}
  {\bibinfo  {journal} {J. Magn. Magn. Mater.}\ }\textbf {\bibinfo {volume}
  {63}},\ \bibinfo {pages} {61 } (\bibinfo {year} {1987})}\BibitemShut
  {NoStop}%
\bibitem [{\citenamefont {Horn}\ \emph {et~al.}(1981)\citenamefont {Horn},
  \citenamefont {Steglich}, \citenamefont {Loewenhaupt}, \citenamefont
  {Scheuer}, \citenamefont {Felsch},\ and\ \citenamefont {Winzer}}]{Horn1981}%
  \BibitemOpen
  \bibfield  {author} {\bibinfo {author} {\bibfnamefont {S.}~\bibnamefont
  {Horn}}, \bibinfo {author} {\bibfnamefont {F.}~\bibnamefont {Steglich}},
  \bibinfo {author} {\bibfnamefont {M.}~\bibnamefont {Loewenhaupt}}, \bibinfo
  {author} {\bibfnamefont {H.}~\bibnamefont {Scheuer}}, \bibinfo {author}
  {\bibfnamefont {W.}~\bibnamefont {Felsch}}, \ and\ \bibinfo {author}
  {\bibfnamefont {K.}~\bibnamefont {Winzer}},\ }\href {\doibase
  10.1007/BF01319546} {\bibfield  {journal} {\bibinfo  {journal} {Zeitschrift
  f{\"u}r Physik B Condens. Matter}\ }\textbf {\bibinfo {volume} {42}},\
  \bibinfo {pages} {125} (\bibinfo {year} {1981})}\BibitemShut {NoStop}%
\bibitem [{\citenamefont {Sera}\ \emph {et~al.}(2001)\citenamefont {Sera},
  \citenamefont {Ichikawa}, \citenamefont {Yokoo}, \citenamefont {Akimitsu},
  \citenamefont {Nishi}, \citenamefont {Kakurai},\ and\ \citenamefont
  {Kunii}}]{PhysRevLett.86.1578}%
  \BibitemOpen
  \bibfield  {author} {\bibinfo {author} {\bibfnamefont {M.}~\bibnamefont
  {Sera}}, \bibinfo {author} {\bibfnamefont {H.}~\bibnamefont {Ichikawa}},
  \bibinfo {author} {\bibfnamefont {T.}~\bibnamefont {Yokoo}}, \bibinfo
  {author} {\bibfnamefont {J.}~\bibnamefont {Akimitsu}}, \bibinfo {author}
  {\bibfnamefont {M.}~\bibnamefont {Nishi}}, \bibinfo {author} {\bibfnamefont
  {K.}~\bibnamefont {Kakurai}}, \ and\ \bibinfo {author} {\bibfnamefont
  {S.}~\bibnamefont {Kunii}},\ }\href {\doibase 10.1103/PhysRevLett.86.1578}
  {\bibfield  {journal} {\bibinfo  {journal} {Phys. Rev. Lett.}\ }\textbf
  {\bibinfo {volume} {86}},\ \bibinfo {pages} {1578} (\bibinfo {year}
  {2001})}\BibitemShut {NoStop}%
\bibitem [{\citenamefont {Zaharko}\ \emph {et~al.}(2003)\citenamefont
  {Zaharko}, \citenamefont {Fischer}, \citenamefont {Schenck}, \citenamefont
  {Kunii}, \citenamefont {Brown}, \citenamefont {Tasset},\ and\ \citenamefont
  {Hansen}}]{PhysRevB.68.214401}%
  \BibitemOpen
  \bibfield  {author} {\bibinfo {author} {\bibfnamefont {O.}~\bibnamefont
  {Zaharko}}, \bibinfo {author} {\bibfnamefont {P.}~\bibnamefont {Fischer}},
  \bibinfo {author} {\bibfnamefont {A.}~\bibnamefont {Schenck}}, \bibinfo
  {author} {\bibfnamefont {S.}~\bibnamefont {Kunii}}, \bibinfo {author}
  {\bibfnamefont {P.-J.}\ \bibnamefont {Brown}}, \bibinfo {author}
  {\bibfnamefont {F.}~\bibnamefont {Tasset}}, \ and\ \bibinfo {author}
  {\bibfnamefont {T.}~\bibnamefont {Hansen}},\ }\href {\doibase
  10.1103/PhysRevB.68.214401} {\bibfield  {journal} {\bibinfo  {journal} {Phys.
  Rev. B}\ }\textbf {\bibinfo {volume} {68}},\ \bibinfo {pages} {214401}
  (\bibinfo {year} {2003})}\BibitemShut {NoStop}%
\bibitem [{\citenamefont {Jang}\ \emph {et~al.}(2014)\citenamefont {Jang},
  \citenamefont {Friemel}, \citenamefont {Ollivier}, \citenamefont {Dukhnenko},
  \citenamefont {Shitsevalova}, \citenamefont {Filipov}, \citenamefont
  {Keimer},\ and\ \citenamefont {Inosov}}]{nmat3976}%
  \BibitemOpen
  \bibfield  {author} {\bibinfo {author} {\bibfnamefont {H.}~\bibnamefont
  {Jang}}, \bibinfo {author} {\bibfnamefont {G.}~\bibnamefont {Friemel}},
  \bibinfo {author} {\bibfnamefont {J.}~\bibnamefont {Ollivier}}, \bibinfo
  {author} {\bibfnamefont {A.~V.}\ \bibnamefont {Dukhnenko}}, \bibinfo {author}
  {\bibfnamefont {N.~Y.}\ \bibnamefont {Shitsevalova}}, \bibinfo {author}
  {\bibfnamefont {V.~B.}\ \bibnamefont {Filipov}}, \bibinfo {author}
  {\bibfnamefont {B.}~\bibnamefont {Keimer}}, \ and\ \bibinfo {author}
  {\bibfnamefont {D.~S.}\ \bibnamefont {Inosov}},\ }\href {\doibase
  http://dx.doi.org/10.1038/nmat3976} {\bibfield  {journal} {\bibinfo
  {journal} {Nat. Mater.}\ }\textbf {\bibinfo {volume} {13}},\ \bibinfo {pages}
  {682} (\bibinfo {year} {2014})}\BibitemShut {NoStop}%
\bibitem [{\citenamefont {Matsumura}\ \emph {et~al.}(2009)\citenamefont
  {Matsumura}, \citenamefont {Yonemura}, \citenamefont {Kunimori},
  \citenamefont {Sera},\ and\ \citenamefont {Iga}}]{PhysRevLett.103.017203}%
  \BibitemOpen
  \bibfield  {author} {\bibinfo {author} {\bibfnamefont {T.}~\bibnamefont
  {Matsumura}}, \bibinfo {author} {\bibfnamefont {T.}~\bibnamefont {Yonemura}},
  \bibinfo {author} {\bibfnamefont {K.}~\bibnamefont {Kunimori}}, \bibinfo
  {author} {\bibfnamefont {M.}~\bibnamefont {Sera}}, \ and\ \bibinfo {author}
  {\bibfnamefont {F.}~\bibnamefont {Iga}},\ }\href {\doibase
  10.1103/PhysRevLett.103.017203} {\bibfield  {journal} {\bibinfo  {journal}
  {Phys. Rev. Lett.}\ }\textbf {\bibinfo {volume} {103}},\ \bibinfo {pages}
  {017203} (\bibinfo {year} {2009})}\BibitemShut {NoStop}%
\bibitem [{\citenamefont {Neupane}\ \emph {et~al.}(2015)\citenamefont
  {Neupane}, \citenamefont {Alidoust}, \citenamefont {Belopolski},
  \citenamefont {Bian}, \citenamefont {Xu}, \citenamefont {Kim}, \citenamefont
  {Shibayev}, \citenamefont {Sanchez}, \citenamefont {Zheng}, \citenamefont
  {Chang}, \citenamefont {Jeng}, \citenamefont {Riseborough}, \citenamefont
  {Lin}, \citenamefont {Bansil}, \citenamefont {Durakiewicz}, \citenamefont
  {Fisk},\ and\ \citenamefont {Hasan}}]{PhysRevB.92.104420}%
  \BibitemOpen
  \bibfield  {author} {\bibinfo {author} {\bibfnamefont {M.}~\bibnamefont
  {Neupane}}, \bibinfo {author} {\bibfnamefont {N.}~\bibnamefont {Alidoust}},
  \bibinfo {author} {\bibfnamefont {I.}~\bibnamefont {Belopolski}}, \bibinfo
  {author} {\bibfnamefont {G.}~\bibnamefont {Bian}}, \bibinfo {author}
  {\bibfnamefont {S.-Y.}\ \bibnamefont {Xu}}, \bibinfo {author} {\bibfnamefont
  {D.-J.}\ \bibnamefont {Kim}}, \bibinfo {author} {\bibfnamefont {P.~P.}\
  \bibnamefont {Shibayev}}, \bibinfo {author} {\bibfnamefont {D.~S.}\
  \bibnamefont {Sanchez}}, \bibinfo {author} {\bibfnamefont {H.}~\bibnamefont
  {Zheng}}, \bibinfo {author} {\bibfnamefont {T.-R.}\ \bibnamefont {Chang}},
  \bibinfo {author} {\bibfnamefont {H.-T.}\ \bibnamefont {Jeng}}, \bibinfo
  {author} {\bibfnamefont {P.~S.}\ \bibnamefont {Riseborough}}, \bibinfo
  {author} {\bibfnamefont {H.}~\bibnamefont {Lin}}, \bibinfo {author}
  {\bibfnamefont {A.}~\bibnamefont {Bansil}}, \bibinfo {author} {\bibfnamefont
  {T.}~\bibnamefont {Durakiewicz}}, \bibinfo {author} {\bibfnamefont
  {Z.}~\bibnamefont {Fisk}}, \ and\ \bibinfo {author} {\bibfnamefont {M.~Z.}\
  \bibnamefont {Hasan}},\ }\href {\doibase 10.1103/PhysRevB.92.104420}
  {\bibfield  {journal} {\bibinfo  {journal} {Phys. Rev. B}\ }\textbf {\bibinfo
  {volume} {92}},\ \bibinfo {pages} {104420} (\bibinfo {year}
  {2015})}\BibitemShut {NoStop}%
\bibitem [{\citenamefont {Kotliar}\ \emph {et~al.}(2006)\citenamefont
  {Kotliar}, \citenamefont {Savrasov}, \citenamefont {Haule}, \citenamefont
  {Oudovenko}, \citenamefont {Parcollet},\ and\ \citenamefont
  {Marianetti}}]{RevModPhys.78.865}%
  \BibitemOpen
  \bibfield  {author} {\bibinfo {author} {\bibfnamefont {G.}~\bibnamefont
  {Kotliar}}, \bibinfo {author} {\bibfnamefont {S.~Y.}\ \bibnamefont
  {Savrasov}}, \bibinfo {author} {\bibfnamefont {K.}~\bibnamefont {Haule}},
  \bibinfo {author} {\bibfnamefont {V.~S.}\ \bibnamefont {Oudovenko}}, \bibinfo
  {author} {\bibfnamefont {O.}~\bibnamefont {Parcollet}}, \ and\ \bibinfo
  {author} {\bibfnamefont {C.~A.}\ \bibnamefont {Marianetti}},\ }\href
  {\doibase 10.1103/RevModPhys.78.865} {\bibfield  {journal} {\bibinfo
  {journal} {Rev. Mod. Phys.}\ }\textbf {\bibinfo {volume} {78}},\ \bibinfo
  {pages} {865} (\bibinfo {year} {2006})}\BibitemShut {NoStop}%
\bibitem [{\citenamefont {Georges}\ \emph {et~al.}(1996)\citenamefont
  {Georges}, \citenamefont {Kotliar}, \citenamefont {Krauth},\ and\
  \citenamefont {Rozenberg}}]{RevModPhys.68.13}%
  \BibitemOpen
  \bibfield  {author} {\bibinfo {author} {\bibfnamefont {A.}~\bibnamefont
  {Georges}}, \bibinfo {author} {\bibfnamefont {G.}~\bibnamefont {Kotliar}},
  \bibinfo {author} {\bibfnamefont {W.}~\bibnamefont {Krauth}}, \ and\ \bibinfo
  {author} {\bibfnamefont {M.~J.}\ \bibnamefont {Rozenberg}},\ }\href {\doibase
  10.1103/RevModPhys.68.13} {\bibfield  {journal} {\bibinfo  {journal} {Rev.
  Mod. Phys.}\ }\textbf {\bibinfo {volume} {68}},\ \bibinfo {pages} {13}
  (\bibinfo {year} {1996})}\BibitemShut {NoStop}%
\bibitem [{\citenamefont {Shim}\ \emph {et~al.}(2007)\citenamefont {Shim},
  \citenamefont {Haule},\ and\ \citenamefont {Kotliar}}]{shim:1615}%
  \BibitemOpen
  \bibfield  {author} {\bibinfo {author} {\bibfnamefont {J.~H.}\ \bibnamefont
  {Shim}}, \bibinfo {author} {\bibfnamefont {K.}~\bibnamefont {Haule}}, \ and\
  \bibinfo {author} {\bibfnamefont {G.}~\bibnamefont {Kotliar}},\ }\href
  {\doibase 10.1126/science.1149064} {\bibfield  {journal} {\bibinfo  {journal}
  {Science}\ }\textbf {\bibinfo {volume} {318}},\ \bibinfo {pages} {1615}
  (\bibinfo {year} {2007})}\BibitemShut {NoStop}%
\bibitem [{\citenamefont {Haule}\ \emph {et~al.}(2010)\citenamefont {Haule},
  \citenamefont {Yee},\ and\ \citenamefont {Kim}}]{PhysRevB.81.195107}%
  \BibitemOpen
  \bibfield  {author} {\bibinfo {author} {\bibfnamefont {K.}~\bibnamefont
  {Haule}}, \bibinfo {author} {\bibfnamefont {C.-H.}\ \bibnamefont {Yee}}, \
  and\ \bibinfo {author} {\bibfnamefont {K.}~\bibnamefont {Kim}},\ }\href
  {\doibase 10.1103/PhysRevB.81.195107} {\bibfield  {journal} {\bibinfo
  {journal} {Phys. Rev. B}\ }\textbf {\bibinfo {volume} {81}},\ \bibinfo
  {pages} {195107} (\bibinfo {year} {2010})}\BibitemShut {NoStop}%
\bibitem [{\citenamefont {Lu}\ and\ \citenamefont
  {Huang}(2016)}]{PhysRevB.94.075132}%
  \BibitemOpen
  \bibfield  {author} {\bibinfo {author} {\bibfnamefont {H.}~\bibnamefont
  {Lu}}\ and\ \bibinfo {author} {\bibfnamefont {L.}~\bibnamefont {Huang}},\
  }\href {\doibase 10.1103/PhysRevB.94.075132} {\bibfield  {journal} {\bibinfo
  {journal} {Phys. Rev. B}\ }\textbf {\bibinfo {volume} {94}},\ \bibinfo
  {pages} {075132} (\bibinfo {year} {2016})}\BibitemShut {NoStop}%
\bibitem [{\citenamefont {Blaha}\ \emph {et~al.}(2001)\citenamefont {Blaha},
  \citenamefont {Schwarz}, \citenamefont {Madsen}, \citenamefont {Kvasnicka},\
  and\ \citenamefont {Luitz}}]{wien2k}%
  \BibitemOpen
  \bibfield  {author} {\bibinfo {author} {\bibfnamefont {P.}~\bibnamefont
  {Blaha}}, \bibinfo {author} {\bibfnamefont {K.}~\bibnamefont {Schwarz}},
  \bibinfo {author} {\bibfnamefont {G.}~\bibnamefont {Madsen}}, \bibinfo
  {author} {\bibfnamefont {D.}~\bibnamefont {Kvasnicka}}, \ and\ \bibinfo
  {author} {\bibfnamefont {J.}~\bibnamefont {Luitz}},\ }\href@noop {} {\emph
  {\bibinfo {title} {{WIEN2k, An Augmented Plane Wave + Local Orbitals Program
  for Calculating Crystal Properties}}}}\ (\bibinfo  {publisher} {Karlheinz
  Schwarz, Techn. Universität Wien, Austria},\ \bibinfo {year}
  {2001})\BibitemShut {NoStop}%
\bibitem [{\citenamefont {Perdew}\ \emph {et~al.}(1996)\citenamefont {Perdew},
  \citenamefont {Burke},\ and\ \citenamefont
  {Ernzerhof}}]{PhysRevLett.77.3865}%
  \BibitemOpen
  \bibfield  {author} {\bibinfo {author} {\bibfnamefont {J.~P.}\ \bibnamefont
  {Perdew}}, \bibinfo {author} {\bibfnamefont {K.}~\bibnamefont {Burke}}, \
  and\ \bibinfo {author} {\bibfnamefont {M.}~\bibnamefont {Ernzerhof}},\ }\href
  {\doibase 10.1103/PhysRevLett.77.3865} {\bibfield  {journal} {\bibinfo
  {journal} {Phys. Rev. Lett.}\ }\textbf {\bibinfo {volume} {77}},\ \bibinfo
  {pages} {3865} (\bibinfo {year} {1996})}\BibitemShut {NoStop}%
\bibitem [{\citenamefont {Fujiwara}\ and\ \citenamefont
  {Korotin}(1999)}]{PhysRevB.59.9903}%
  \BibitemOpen
  \bibfield  {author} {\bibinfo {author} {\bibfnamefont {T.}~\bibnamefont
  {Fujiwara}}\ and\ \bibinfo {author} {\bibfnamefont {M.}~\bibnamefont
  {Korotin}},\ }\href {\doibase 10.1103/PhysRevB.59.9903} {\bibfield  {journal}
  {\bibinfo  {journal} {Phys. Rev. B}\ }\textbf {\bibinfo {volume} {59}},\
  \bibinfo {pages} {9903} (\bibinfo {year} {1999})}\BibitemShut {NoStop}%
\bibitem [{\citenamefont {Anisimov}\ \emph {et~al.}(1997)\citenamefont
  {Anisimov}, \citenamefont {Aryasetiawan},\ and\ \citenamefont
  {Lichtenstein}}]{jpcm:1997}%
  \BibitemOpen
  \bibfield  {author} {\bibinfo {author} {\bibfnamefont {V.~I.}\ \bibnamefont
  {Anisimov}}, \bibinfo {author} {\bibfnamefont {F.}~\bibnamefont
  {Aryasetiawan}}, \ and\ \bibinfo {author} {\bibfnamefont {A.~I.}\
  \bibnamefont {Lichtenstein}},\ }\href {\doibase 10.1088/0953-8984/9/4/002}
  {\bibfield  {journal} {\bibinfo  {journal} {J. Phys.: Condens. Matter}\
  }\textbf {\bibinfo {volume} {9}},\ \bibinfo {pages} {767} (\bibinfo {year}
  {1997})}\BibitemShut {NoStop}%
\bibitem [{\citenamefont {Gull}\ \emph {et~al.}(2011)\citenamefont {Gull},
  \citenamefont {Millis}, \citenamefont {Lichtenstein}, \citenamefont
  {Rubtsov}, \citenamefont {Troyer},\ and\ \citenamefont
  {Werner}}]{RevModPhys.83.349}%
  \BibitemOpen
  \bibfield  {author} {\bibinfo {author} {\bibfnamefont {E.}~\bibnamefont
  {Gull}}, \bibinfo {author} {\bibfnamefont {A.~J.}\ \bibnamefont {Millis}},
  \bibinfo {author} {\bibfnamefont {A.~I.}\ \bibnamefont {Lichtenstein}},
  \bibinfo {author} {\bibfnamefont {A.~N.}\ \bibnamefont {Rubtsov}}, \bibinfo
  {author} {\bibfnamefont {M.}~\bibnamefont {Troyer}}, \ and\ \bibinfo {author}
  {\bibfnamefont {P.}~\bibnamefont {Werner}},\ }\href {\doibase
  10.1103/RevModPhys.83.349} {\bibfield  {journal} {\bibinfo  {journal} {Rev.
  Mod. Phys.}\ }\textbf {\bibinfo {volume} {83}},\ \bibinfo {pages} {349}
  (\bibinfo {year} {2011})}\BibitemShut {NoStop}%
\bibitem [{\citenamefont {Werner}\ \emph {et~al.}(2006)\citenamefont {Werner},
  \citenamefont {Comanac}, \citenamefont {de' Medici}, \citenamefont {Troyer},\
  and\ \citenamefont {Millis}}]{PhysRevLett.97.076405}%
  \BibitemOpen
  \bibfield  {author} {\bibinfo {author} {\bibfnamefont {P.}~\bibnamefont
  {Werner}}, \bibinfo {author} {\bibfnamefont {A.}~\bibnamefont {Comanac}},
  \bibinfo {author} {\bibfnamefont {L.}~\bibnamefont {de' Medici}}, \bibinfo
  {author} {\bibfnamefont {M.}~\bibnamefont {Troyer}}, \ and\ \bibinfo {author}
  {\bibfnamefont {A.~J.}\ \bibnamefont {Millis}},\ }\href {\doibase
  10.1103/PhysRevLett.97.076405} {\bibfield  {journal} {\bibinfo  {journal}
  {Phys. Rev. Lett.}\ }\textbf {\bibinfo {volume} {97}},\ \bibinfo {pages}
  {076405} (\bibinfo {year} {2006})}\BibitemShut {NoStop}%
\bibitem [{\citenamefont {Haule}(2007)}]{PhysRevB.75.155113}%
  \BibitemOpen
  \bibfield  {author} {\bibinfo {author} {\bibfnamefont {K.}~\bibnamefont
  {Haule}},\ }\href {\doibase 10.1103/PhysRevB.75.155113} {\bibfield  {journal}
  {\bibinfo  {journal} {Phys. Rev. B}\ }\textbf {\bibinfo {volume} {75}},\
  \bibinfo {pages} {155113} (\bibinfo {year} {2007})}\BibitemShut {NoStop}%
\bibitem [{\citenamefont {S\'emon}\ \emph {et~al.}(2014)\citenamefont
  {S\'emon}, \citenamefont {Yee}, \citenamefont {Haule},\ and\ \citenamefont
  {Tremblay}}]{PhysRevB.90.075149}%
  \BibitemOpen
  \bibfield  {author} {\bibinfo {author} {\bibfnamefont {P.}~\bibnamefont
  {S\'emon}}, \bibinfo {author} {\bibfnamefont {C.-H.}\ \bibnamefont {Yee}},
  \bibinfo {author} {\bibfnamefont {K.}~\bibnamefont {Haule}}, \ and\ \bibinfo
  {author} {\bibfnamefont {A.-M.~S.}\ \bibnamefont {Tremblay}},\ }\href
  {\doibase 10.1103/PhysRevB.90.075149} {\bibfield  {journal} {\bibinfo
  {journal} {Phys. Rev. B}\ }\textbf {\bibinfo {volume} {90}},\ \bibinfo
  {pages} {075149} (\bibinfo {year} {2014})}\BibitemShut {NoStop}%
\bibitem [{\citenamefont {G\"urel}\ and\ \citenamefont
  {Eryi\ifmmode~\breve{g}\else \u{g}\fi{}it}(2010)}]{PhysRevB.82.104302}%
  \BibitemOpen
  \bibfield  {author} {\bibinfo {author} {\bibfnamefont {T.}~\bibnamefont
  {G\"urel}}\ and\ \bibinfo {author} {\bibfnamefont {R.}~\bibnamefont
  {Eryi\ifmmode~\breve{g}\else \u{g}\fi{}it}},\ }\href {\doibase
  10.1103/PhysRevB.82.104302} {\bibfield  {journal} {\bibinfo  {journal} {Phys.
  Rev. B}\ }\textbf {\bibinfo {volume} {82}},\ \bibinfo {pages} {104302}
  (\bibinfo {year} {2010})}\BibitemShut {NoStop}%
\bibitem [{\citenamefont {Nakamura}\ \emph {et~al.}(1994)\citenamefont
  {Nakamura}, \citenamefont {Goto}, \citenamefont {Kunii}, \citenamefont
  {Iwashita},\ and\ \citenamefont {Tamaki}}]{JPSJ.63.623}%
  \BibitemOpen
  \bibfield  {author} {\bibinfo {author} {\bibfnamefont {S.}~\bibnamefont
  {Nakamura}}, \bibinfo {author} {\bibfnamefont {T.}~\bibnamefont {Goto}},
  \bibinfo {author} {\bibfnamefont {S.}~\bibnamefont {Kunii}}, \bibinfo
  {author} {\bibfnamefont {K.}~\bibnamefont {Iwashita}}, \ and\ \bibinfo
  {author} {\bibfnamefont {A.}~\bibnamefont {Tamaki}},\ }\href {\doibase
  10.1143/JPSJ.63.623} {\bibfield  {journal} {\bibinfo  {journal} {J. Phys.
  Soc. Jpn.}\ }\textbf {\bibinfo {volume} {63}},\ \bibinfo {pages} {623}
  (\bibinfo {year} {1994})}\BibitemShut {NoStop}%
\bibitem [{\citenamefont {Leger}\ \emph {et~al.}(1985)\citenamefont {Leger},
  \citenamefont {Rossat-Mignod}, \citenamefont {Kunii},\ and\ \citenamefont
  {Kasuya}}]{LEGER1985995}%
  \BibitemOpen
  \bibfield  {author} {\bibinfo {author} {\bibfnamefont {J.}~\bibnamefont
  {Leger}}, \bibinfo {author} {\bibfnamefont {J.}~\bibnamefont
  {Rossat-Mignod}}, \bibinfo {author} {\bibfnamefont {S.}~\bibnamefont
  {Kunii}}, \ and\ \bibinfo {author} {\bibfnamefont {T.}~\bibnamefont
  {Kasuya}},\ }\href {\doibase http://dx.doi.org/10.1016/0038-1098(85)90172-3}
  {\bibfield  {journal} {\bibinfo  {journal} {Solid State Commun.}\ }\textbf
  {\bibinfo {volume} {54}},\ \bibinfo {pages} {995 } (\bibinfo {year}
  {1985})}\BibitemShut {NoStop}%
\bibitem [{\citenamefont {L{\"u}thi}\ \emph {et~al.}(1984)\citenamefont
  {L{\"u}thi}, \citenamefont {Blumenr{\"o}der}, \citenamefont {Hillebrands},
  \citenamefont {Zirngiebl}, \citenamefont {G{\"u}ntherodt},\ and\
  \citenamefont {Winzer}}]{Lüthi1984}%
  \BibitemOpen
  \bibfield  {author} {\bibinfo {author} {\bibfnamefont {B.}~\bibnamefont
  {L{\"u}thi}}, \bibinfo {author} {\bibfnamefont {S.}~\bibnamefont
  {Blumenr{\"o}der}}, \bibinfo {author} {\bibfnamefont {B.}~\bibnamefont
  {Hillebrands}}, \bibinfo {author} {\bibfnamefont {E.}~\bibnamefont
  {Zirngiebl}}, \bibinfo {author} {\bibfnamefont {G.}~\bibnamefont
  {G{\"u}ntherodt}}, \ and\ \bibinfo {author} {\bibfnamefont {K.}~\bibnamefont
  {Winzer}},\ }\href {\doibase 10.1007/BF01469434} {\bibfield  {journal}
  {\bibinfo  {journal} {Zeitschrift f{\"u}r Physik B Condens. Matter}\ }\textbf
  {\bibinfo {volume} {58}},\ \bibinfo {pages} {31} (\bibinfo {year}
  {1984})}\BibitemShut {NoStop}%
\bibitem [{\citenamefont {Sato}(1985)}]{SATO1985310}%
  \BibitemOpen
  \bibfield  {author} {\bibinfo {author} {\bibfnamefont {S.}~\bibnamefont
  {Sato}},\ }\href {\doibase http://dx.doi.org/10.1016/0304-8853(85)90288-4}
  {\bibfield  {journal} {\bibinfo  {journal} {J. Magn. Magn. Mater.}\ }\textbf
  {\bibinfo {volume} {52}},\ \bibinfo {pages} {310 } (\bibinfo {year}
  {1985})}\BibitemShut {NoStop}%
\bibitem [{val()}]{valence}%
  \BibitemOpen
  \href@noop {} {}\bibinfo {note} {In Ce-based heavy fermion compounds, the
  Ce-4$f$ electrons usually exhibit fascinating mixed-valence (or equivalently
  valence fluctuation) behaviors, resulting in non-integer $f$ occupancy.
  However, the valence fluctuation in Kondo compound CeB$_{6}$ is relatively
  small.}\BibitemShut {Stop}%
\bibitem [{\citenamefont {Haule}\ and\ \citenamefont
  {Birol}(2015)}]{PhysRevLett.115.256402}%
  \BibitemOpen
  \bibfield  {author} {\bibinfo {author} {\bibfnamefont {K.}~\bibnamefont
  {Haule}}\ and\ \bibinfo {author} {\bibfnamefont {T.}~\bibnamefont {Birol}},\
  }\href {\doibase 10.1103/PhysRevLett.115.256402} {\bibfield  {journal}
  {\bibinfo  {journal} {Phys. Rev. Lett.}\ }\textbf {\bibinfo {volume} {115}},\
  \bibinfo {pages} {256402} (\bibinfo {year} {2015})}\BibitemShut {NoStop}%
\bibitem [{\citenamefont {Flint}\ \emph {et~al.}(2014)\citenamefont {Flint},
  \citenamefont {Chandra},\ and\ \citenamefont {Coleman}}]{rebecca:2014}%
  \BibitemOpen
  \bibfield  {author} {\bibinfo {author} {\bibfnamefont {R.}~\bibnamefont
  {Flint}}, \bibinfo {author} {\bibfnamefont {P.}~\bibnamefont {Chandra}}, \
  and\ \bibinfo {author} {\bibfnamefont {P.}~\bibnamefont {Coleman}},\ }\href
  {\doibase 10.7566/JPSJ.83.061003} {\bibfield  {journal} {\bibinfo  {journal}
  {J. Phys. Soc. Japan}\ }\textbf {\bibinfo {volume} {83}},\ \bibinfo {pages}
  {061003} (\bibinfo {year} {2014})}\BibitemShut {NoStop}%
\bibitem [{\citenamefont {Mydosh}\ and\ \citenamefont
  {Oppeneer}(2011)}]{RevModPhys.83.1301}%
  \BibitemOpen
  \bibfield  {author} {\bibinfo {author} {\bibfnamefont {J.~A.}\ \bibnamefont
  {Mydosh}}\ and\ \bibinfo {author} {\bibfnamefont {P.~M.}\ \bibnamefont
  {Oppeneer}},\ }\href {\doibase 10.1103/RevModPhys.83.1301} {\bibfield
  {journal} {\bibinfo  {journal} {Rev. Mod. Phys.}\ }\textbf {\bibinfo {volume}
  {83}},\ \bibinfo {pages} {1301} (\bibinfo {year} {2011})}\BibitemShut
  {NoStop}%
\end{thebibliography}%

\end{document}